\documentclass[12pt]{article}
\usepackage{multirow}
\usepackage{supertabular}
\usepackage{amsmath}
\usepackage{amssymb}
\input{epsf}
\begin{document}

\begin{center}
{\Large\bf Non-yrast nuclear spectra in a model of coherent
quadrupole-octupole motion}\\
\bigskip\bigskip

N. Minkov, S. Drenska, \\
\smallskip
{\em Institute of Nuclear Research and Nuclear Energy, Bulgarian
Academy of Sciences,
Tzarigrad Road 72, BG-1784 Sofia, Bulgaria} \\
\medskip

M. Strecker, W. Scheid and H. Lenske\\
\smallskip
{\em Institut f\"{u}r Theoretische Physik der
Justus-Liebig-Universit\"at, Heinrich-Buff-Ring 16, D--35392
Giessen, Germany}
\end{center}
\bigskip

\begin{abstract}

A model assuming coherent quadrupole-octupole vibrations and rotations is
applied to describe non-yrast energy sequences with alternating parity in
several even-even nuclei from different regions, namely $^{152,154}$Sm,
$^{154,156,158}$Gd, $^{236}$U and $^{100}$Mo. Within the model scheme the yrast
alternating-parity band is composed by the members of the ground-state band and
the lowest negative-parity levels with odd angular momenta. The non-yrast
alternating-parity sequences unite levels of $\beta$-bands with higher
negative-parity levels. The model description reproduces the structure of the
considered alternating-parity spectra together with the observed B(E1), B(E2)
and B(E3) transition probabilities within and between the different
level-sequences. B(E1) and B(E3) reduced probabilities for transitions
connecting states with opposite parity in the non-yrast alternating-parity
bands are predicted. The implemented study outlines the limits of the
considered band-coupling scheme and provides estimations about the collective
energy potential which governs the quadrupole-octupole properties of the
considered nuclei.

\end{abstract}

PACS: 21.60.Ev, 21.10.Re, 27.70.+q, 27.90.+b, 27.60.+j
\newpage

\section{Introduction}

A typical manifestation of the reflection-asymmetric quadrupole-octupole
deformation in the energy spectra of even-even atomic nuclei is the formation
of level sequences with alternating parities \cite{BN96}. Usually the levels
with opposite parity are related through enhanced electric E1 and/or E3
transitions. The negative-parity sequence is shifted up with respect to the
positive-parity sequence due to a tunneling of the system between the two
opposite orientations along the principal symmetry axis. The magnitude of the
energy shift corresponds to the softness of the shape with respect to the
octupole deformation. The typical alternating-parity band is formed by the
members of the ground-state ($g$) band and the levels of the lowest
negative-parity sequence with odd angular momenta. In the relatively narrow
region of the light actinide nuclei Rn, Ra and Th these two sequences merge
into a single rotation band also called ``octupole band''
\cite{Th226,Th224,Cocks97}. The octupole band develops in the higher angular
momenta and indicates the appearance of a quite stiff octupole deformation.
Away from the light actinide region both sequences diverge and do not form a
single rotation band in the conventional meaning. Nevertheless, in some heavier
actinides, like U and Pu and some rare-earth isotopes like Nd, Sm, Gd and Dy,
they still remain related by E1 and E3  transitions, which indicates the
presence of a soft octupole mode in the collective motion. In this case the
term ``alternating-parity band (or spectrum)'' does not have the same strict
meaning as in the light actinide nuclei but simply refers to sequences of
levels with opposite parities which could be connected (coupled) through
electric transitions.

Various theoretical models have been developed over the years to explain and
describe the formation of alternating-parity (or octupole) bands in the stiff
and soft octupole regimes of coupling between the $g$-band and the lowest
negative-parity sequences in different nuclear regions
\cite{Krappe69}--\cite{Buck08}. Particularly, a collective model assuming
coherent quadrupole-octupole vibrations and rotations \cite{b2b3mod} was
applied to the nuclei $^{150}$Nd, $^{152}$Sm, $^{154}$Gd and $^{156}$Dy with
the presence of a soft octupole collectivity. Although the $g$-band and the
lowest negative-parity bands in these nuclei were successfully described as
members of an yrast alternating-parity band together with the attendant B(E1)
and B(E2) transition probabilities, a question arises about the validity of
such a consideration with respect to the higher-energy (non-yrast) part of the
spectrum.

The purpose of the present work is to clarify the above question within the
model of Coherent Quadrupole-Octupole Motion  (CQOM) \cite{b2b3mod} by
examining the possible formation of non-yrast alternating-parity structures in
addition to the yrast band.  For this reason the model scheme is extended by
assuming that the excited $\beta$-bands can be connected to higher
negative-parity sequences with odd angular momenta. Therefore, it is supposed
that the quadrupole-octupole structure of the spectrum develops along the
non-yrast regions of the energy spectrum. Such a study provides not only a test
of the model in the higher energy parts of the spectra, but also gives an
interpretation of a larger number of data that may guide the experimental
search for similar level structures in other nuclear regions. In principle, the
systematic analysis of the non-yrast levels with alternating parity may favour
different band-coupling schemes in the different nuclear regions allowing one
to compare the capabilities of various theoretical models. For example, an
extended study of non-yrast energy sequences with different parities has been
implemented within the Extended Coherent States Model \cite{RRR06} by
considering a coupling of the $\beta$ and $\gamma$ bands with respective bands
possessing the same spins but opposite parities, as well as a coupling between
$K^{\pi}=1^{+}$ and $K^{\pi}=1^{-}$ energy sequences. In the model scheme of
the present work the positive-parity $\beta$-band appears connected to a
negative-parity non-yrast sequence with odd angular momenta in the same way as
in the yrast alternating-parity configuration. This is a consequence of the
assumed mechanism of coupling between the quadrupole and octupole vibration
modes. Therefore, the present work suggests a different band-coupling scheme
and supposes a persistent role of the quadrupole-octupole motion in the forming
of the higher-energy (non-yrast) part of the spectrum. Of course, by developing
such an approach one should keep in mind the non-conventional meaning of the
term ``alternating-parity band'' mentioned above. Also, presently the CQOM
model is limited to excitations associated with the axial quadrupole and
octupole degrees of freedom. Therefore, the study is focused on the related
part of the collective spectrum, while other kinds of excitation modes as the
$\gamma$-vibrations remain beyond the present consideration.

The paper is organized as follows. In Sec. 2 the CQOM model is presented and
the model mechanism for the appearance of non-yrast alternating parity bands is
shown. Model expressions for reduced B(E1), B(E2) and B(E3) transitions in the
non-yrast spectra are given in Sec. 3. In Sec. 4 numerical results and
discussion on the application of the model to the nuclei of different regions
are given. Sec. 5 contains concluding remarks.

\section{Model of Coherent Quadrupole--Octupole Motion}

The CQOM model \cite{b2b3mod} is a particular realization of the more general
geometric concept of collective nuclear motion characterized by the
quadrupole-octupole shape deformations \cite{BN96}. The expansion of the
surface radius $R(\theta,\varphi)$, in polar coordinates, with respect to
spherical harmonics up to multipolarity $\lambda =3$ is given by
\begin{eqnarray}
R(\theta ,\varphi)=R_{0}\left[ 1+\sum_{\lambda
=2}^{3}\sum^{\lambda}_{\mu = -\lambda}\alpha_{\lambda \mu}
Y_{\lambda\mu}^{\ast}(\theta,\varphi)\right],
\end{eqnarray}
where $R_0$ is the spherical radius and $\alpha_{\lambda \mu}$ are the twelve
quadrupole and octupole collective coordinates in the laboratory frame. The
collective coordinates are transformed into a body-fixed frame
\begin{equation}
a_{\lambda\nu}=\sum_{\mu}\alpha_{\lambda\mu}D^{\lambda}_{\mu\nu}(\hat{\theta}),
\end{equation}
determined by the ``canonical'' quadrupole coordinates $a_0=a_{20}$ and
$a_2=a_{22}=a_{2-2}$ and the three Euler angles
$\hat{\theta}=(\theta_{1},\theta_{2},\theta_{3})$. The remaining seven octupole
coordinates $a_{3\mu}$ ($\mu =-3,...,3$) together with $a_0$ and $a_2$
determine the quadrupole-octupole shape of the nucleus. In the particular case
of axial symmetry the quadrupole-octupole deformation represents a pear-like
shape determined by the only non-zero coordinates $\beta_2\equiv a_0$ and
$\beta_3\equiv a_{30}$. The respective physical states of the nucleus in the
intrinsic (body-fixed) frame are characterized by the symmetrization group
D$_{\infty}$ which consists of arbitrary (infinite number) rotations about the
intrinsic $z$-axis and rotations about the axes perpendicular to $z$ through
the angle $\pi$. In principal the symmetrization group of the nucleus in the
intrinsic frame is determined by the rotations $g$ satisfying a set of
equations in the form $D^{\lambda}_{\mu\nu}(g)=0$, which in the case of axial
symmetry is $D^{\lambda}_{\mu 0}(g)=0$ for all $\mu \neq 0$ \cite{GSDD11}.

In the CQOM model \cite{b2b3mod} the geometric concept is implemented in the
limits of the axial symmetry. It is considered that the even--even nucleus can
oscillate with respect to the quadrupole $\beta_2$ and octupole $\beta_3$ axial
deformation variables, which are mixed through a centrifugal
(rotation-vibration) interaction. The collective Hamiltonian of the nucleus is
then taken in the form
\begin{eqnarray}
H_{qo}&=&-\frac{\hbar^2}{2B_2}\frac{\partial^2}{\partial\beta_2^2}
-\frac{\hbar^2}{2B_3}\frac{\partial^2}{\partial\beta_3^2}+
U(\beta_2,\beta_3,I) \ , \label{Hqo}
\end{eqnarray}
where
\begin{equation}
U(\beta_2,\beta_3, I)=\frac{1}{2}C_2{\beta_2}^{2}+
\frac{1}{2}C_3{\beta_3}^{2} +
\frac{X(I)}{d_2\beta_2^2+d_3\beta_3^2}\ ,
\label{Ub2b3I}
\end{equation}
with $X(I)=[d_0+I(I+1)]/2$.  $B_2$ and $B_3$ are effective quadrupole and
octupole mass parameters and $C_2$ and $C_3$ are stiffness parameters for the
respective oscillation modes. The quantity ${\mathcal{J}}^{(\mbox{\scriptsize
quad+oct})}=(d_2\beta_2^2+d_3\beta_3^2)$ can be associated to the moment of
inertia of an axially symmetric quadrupole-octupole deformed shape
\cite{JPDav68} with $d_2$ and $d_3$ being inertia parameters. The energy
potential (\ref{Ub2b3I}) represents a two-dimensional surface determined by the
variables $\beta_2$ and $\beta_3$ with an angular-momentum-dependent repulsive
core at zero deformation (see Fig. 1 in \cite{b2b3mod}). The parameter $d_0$ in
the centrifugal factor $X(I)$ characterizes the repulsive core at $I=0$ and
determines the overall energy scale for the rotation part of the energy.

The model Hamiltonian (\ref{Hqo}) represents a D$_{\infty}$ invariant. Also, it
is important to remark that (\ref{Hqo}) corresponds to a class of
axial-symmetric Hamiltonians  \cite{DD93}, \cite{DD95}, \cite{AQOA} whose
kinetic vibration parts are derived by ignoring the non-axial degrees of
freedom (e.g. $\gamma$-vibrations) in a way similar to the approach of Davidov
and Chaban \cite{DCh60}. The scalar product in the space of the wave functions
(e.g. see Eqs. (2) and (4) in \cite{AQOA}) corresponding to the particular form
of the $\beta_2$- and $\beta_3$-derivatives in (\ref{Hqo}) is characterized by
a unit weight factor, i.e. $\langle\Phi_2|\Phi_1\rangle=\int\int
d\beta_2d\beta_3 \Phi_2^\ast (\beta_2 ,\beta_3) \Phi_1(\beta_2 ,\beta_3)$.

If a condition for the simultaneous presence of nonzero coordinates
$(\beta_{2}^{\mbox{\scriptsize min}}, \beta_{3}^{\mbox{\scriptsize min}})$ of
the potential minimum is imposed, the stiffness and inertial parameters are
correlated as $d_2/C_2=d_3/C_3$ (see Eqs. (3)--(6) in \cite{b2b3mod}). In this
case the potential bottom represents an ellipse in the space of $\beta_2$ and
$\beta_3$ which surrounds the infinite zero-deformation core (see Fig. 3 in
\cite{b2b3mod}). If prolate quadrupole deformations $\beta_2>0$ are considered,
the system is characterized by oscillations between positive and negative
$\beta_3$-values along the ellipse surrounding the potential core. By
introducing polar-type of curvilinear or, more precise, ellipsoidal variables
\begin{eqnarray} \eta=\left[
\frac{2(d_2\beta_2^2+d_3\beta_3^2)}{d_2+d_3}\right]^{\frac{1}{2}} \qquad
\mbox{and}\qquad \phi=\arctan\left( {\frac{\beta_3}{\beta_2}
\sqrt{\frac{d_3}{d_2}}}\right )\ , \nonumber
\end{eqnarray}
such that
\begin{eqnarray}
\beta_{2}=p\eta\cos\phi, \qquad \beta_{3}=q\eta\sin\phi,
\label{polar}
\end{eqnarray}
with
\begin{eqnarray}
p=\sqrt{\frac{d}{d_{2}}},\qquad q=\sqrt{\frac{d}{d_{3}}}\qquad \mathrm{and}
\qquad d=\frac{1}{2}(d_{2}+d_{3}),
\label{pqd}
\end{eqnarray}
the potential (\ref{Ub2b3I}) appears in the form
\begin{eqnarray}
U_{I}(\eta)=\frac{1}{2}C\eta^2+\frac{X(I)}{d\eta^2}\ ,
\label{cpotenc}
\end{eqnarray}
where $C=(d/d_2)C_2=(d/d_3)C_3$.

Further, it is assumed that the quadrupole  and octupole modes are represented
in the collective motion with the same oscillation frequencies
$\omega_{2}=\omega_{3}=\omega$, with
\begin{eqnarray}
\omega=\sqrt{\frac{C_2}{B_2}}=\sqrt{\frac{C_3}{B_3}}\equiv \sqrt{\frac{C}{B}}.
\label{om}
\end{eqnarray}
The condition (\ref{om}) imposes certain correlations between the mass,
stiffness and inertia parameters of the model Hamiltonian (\ref{Hqo}),
corresponding to a coherent quadrupole-octupole motion of the system. Note that
here the term ``coherent'' is used in the context of the mixing between the
quadrupole and octupole degrees of freedom, which is different from the meaning
of the same term used in \cite{RRR06}. In this case the Hamiltonian is obtained
in a simple form
\begin{eqnarray}
H_{qo}&=&-\frac{\hbar^2}{2B}\left[\frac{\partial^2}{\partial\eta^2}+
\frac{1}{\eta }\frac{\partial}{\partial \eta }+ \frac{1}{\eta
^2}\frac{\partial^2}{\partial \phi^2} \right] +U_{I}(\eta
) \ . \label{Hqo02}
\end{eqnarray}
It allows an exact separation of variables in the wave function $\Phi(\eta
,\phi)=\psi(\eta )\varphi(\phi)$ with the subsequent equations for $\psi(\eta
)$ and $\varphi(\phi)$
\begin{eqnarray}
\frac{\partial^2}{\partial \eta ^2}\psi(\eta )+ \frac{1}{\eta
}\frac{\partial}{\partial \eta }\psi(\eta )
+\frac{2B}{\hbar^2}\left[ E-\frac{\hbar^2} {2B}\frac{k^2}{\eta
^2}-U_{I}(\eta )\right] \psi (\eta )&=&0 \ ; \label{Hqo03} \\
\frac{\partial^2}{\partial
\phi^2}\varphi(\phi)+k^2\varphi(\phi)&=&0 \ , \label{wphi}
\end{eqnarray}
where $k$ is the separation quantum number. Eq. (\ref{Hqo03}) with the
potential (\ref{cpotenc}) is similar to the equation for the Davidson potential
\cite{Dav32} and has the following analytic solution for the energy spectrum
\cite{b2b3mod}
\begin{equation}
E_{n,k}(I) =\hbar\omega \left[ 2n+1+\sqrt{k^2+bX(I)}\right],
\label{enspect}
\end{equation}
where $\omega$ is defined in (\ref{om}), $n=0,1,2,...$ and $b=2B/(\hbar^2 d)$.
The quantum numbers $n$ and $k$ have the meaning of ``radial'' and ``angular''
oscillation quantum numbers, respectively. The normalized ``radial''
eigenfunctions $\psi(\eta )$  are obtained in terms of the generalized Laguerre
polynomials
\begin{equation}
\psi^I_{n,k}(\eta )=\sqrt
{\frac {2c\Gamma(n+1)}{\Gamma(n+2s+1)}}
e^{-c\eta^2/2}(c\eta^{2})^sL^{2s}_n(c\eta^2)\ , \label{psieta1}
\end{equation}
with $c=\sqrt{BC}/\hbar$ and $s=(1/2)\sqrt{k^2+bX(I)}$. Eq.~(\ref{wphi}) in the
``angular'' variable $\phi$ is solved under the boundary condition
$\varphi(-\pi /2)=\varphi(\pi /2)=0$. This is equivalent to the consideration
of an infinite potential wall at $\beta_2=0$ (or $\varphi =\pm\pi /2$). Then
one has two identical solutions for $\beta_2>0$ and $\beta_2<0$. As mentioned
above the physical space of the model is taken in the prolate $\beta_2>0$ half
of the $(\beta_2,\beta_3)$-plane. (See Figs. 4 and 5 in \cite{b2b3mod} and the
related text in that reference). Within this half-plane Eq.~(\ref{wphi}) has
two different solutions with positive and negative parities, $\pi =(+)$ and
$\pi=(-)$, respectively
\begin{eqnarray}
\varphi_{k}^{+}(\phi)&=& \sqrt{2/\pi}\cos (k\phi ) \ , \qquad k=1, 3, 5,
...\ ,\label{parplus} \\
\varphi_{k}^{-}(\phi)&=& \sqrt{2/\pi}\sin (k\phi ) \ , \qquad k=2, 4, 6,
...\ . \label{parminus}
\end{eqnarray}
Note that the square root term in the wave function $\psi^I_{n,k}(\eta )$,
Eq.~(\ref{psieta1}), differs from the respective term used in Eq.~(24) of
Ref.~\cite{b2b3mod} by the factor $c$ which is newly included in the numerator.
(In \cite{b2b3mod} the quantity $c$ is denoted by `$a$' which in the case of
odd nuclei leads to confusion with the notation for the decoupling factor.) One
can easily check that this factor is necessary to normalize $\psi^I_{n,k}(\eta
)$ to unity. The results for the transition probabilities obtained in
\cite{b2b3mod} are not affected by the missing factor $c$ due to the use of
overall scaling factors in Eqs.~(46) and (47) of \cite{b2b3mod}.

Since the consideration is restricted to axial deformations only, the
projection  $K$ of the collective angular momentum on the principal symmetry
axis is taken zero. Then the total wave function of the coherent
quadrupole-octuple vibration and collective rotation of an even-even nucleus
has the form
\begin{eqnarray}
\Psi^{\pi}_{nkIM0}(\eta ,\phi)= \sqrt{\frac{2I+1}{8\pi^2}} D^{I}_{M\,0}(\theta )
\Phi^{\pi}_{nkI} (\eta,\phi)
\ , \label{wftot}
\end{eqnarray}
where $D^{I}_{M\,0}(\theta )$ is the Wigner function defined according to the
phase convention in \cite{BM75}. Note that due to a different phase convention
in some other works, e.g. in \cite{EG87} and \cite{RS80}, the complex
conjugated $D$-function appears in the rotation part. The relations between the
different definitions of the $D$-function are given in Table 4.2 in
\cite{VMK88}. The quadrupole-octupole vibration part of (\ref{wftot}) is
\begin{eqnarray}
\Phi^{\pi}_{nkI} (\eta,\phi)= \psi_{nk}^{I}(\eta )\varphi^{\pi}_{k}(\phi)\ .
\label{wvib}
\end{eqnarray}
The quantum numbers of the quadrupole-octupole vibration function (\ref{wvib})
are determined by the requirement for a conservation of the
$\mathcal{R}\mathcal{P}$-symmetry of the total wave function (\ref{wftot}).
($\mathcal{P}$ is the parity operator and $\mathcal{R}$ represents a rotation
by an angle $\pi$ about an axis perpendicular to the intrinsic $z$-axis) The
$\mathcal{R}$-symmetry of the rotation function $D^{I}_{M\,0}(\theta )$ is
characterized by the factor $(-1)^I$, while the action of $\mathcal{P}$ on
$\Phi^{\pi}_{nkI} (\eta,\phi)$ gives the factor $\pi =\pm$. Then the
conservation of the $\mathcal{R}\mathcal{P}$-symmetry is equivalent to the
conservation of the so called simplex quantum number $simplex=\pi (-1)^I=1$.
This condition imposes a positive parity for the states with even angular
momentum, and negative parity for the odd angular momentum states, i.e. one has
\begin{eqnarray}
\Phi^{+}_{nkI} (\eta,\phi)&=& \psi_{nk}^{I}(\eta )\varphi^{+}_{k}(\phi)\
\mbox{for}\ I=\mbox{even}\nonumber\\
\Phi^{-}_{nkI} (\eta,\phi)&=& \psi_{nk}^{I}(\eta )\varphi^{-}_{k}(\phi)\
\mbox{for}\ I=\mbox{odd}\nonumber .
\end{eqnarray}
It should be noted that the above conditions are in conjunction with the
transformation properties of the variables $\eta$ and $\phi$ in (\ref{wvib})
under the rotation $\mathcal{R}$ ($\eta$ is invariant, while $\phi$ changes in
sign) so that together with the simplex conservation condition the total wave
function (\ref{wftot}) appears to be an D$_{\infty}$ invariant as it should be
due to the axial symmetry.

The structure of the energy spectrum is determined by the oscillator quantum
numbers $n$  (``radial'') and $k$ (``angular'')  in Eq.~(\ref{enspect}). Since
according to Eqs.~(\ref{parplus}) and (\ref{parminus}) $k$ obtains different
values for the states with opposite parity the energy sequences with even and
odd angular momenta corresponding to a given $n$ appear shifted to each other,
i.e. a parity shift effect is observed. In \cite{b2b3mod} it was supposed that
the $g$-band and the lowest negative-parity band belong to a $n=0$ set with
$k=k^{(+)}=1$ for $g$ and $k=k^{(-)}=2$ for the negative-parity band. In the
present work the model scheme is extended through the following three
suppositions.

i) The energy spectrum determined by the coherent axial quadrupole-octupole
vibrations and rotations consists of couples of level-sequences with opposite
parity. The sequences in each couple are characterized by the same value of the
quantum number $n=0,1,2,...$ and by different values of $k$,
$k=k_n^{(+)}=1\,\mbox{or}\,3\,\mbox{or}\,5\,\mbox{or}\,...$ for the even-$I$
sequence, and $k=k_n^{(-)}=2\,\mbox{or}\,4\,\mbox{or}\,6\,\mbox{or}\,...$ for
the odd-$I$ sequence.

ii) The lowest values of the ``radial'' quantum number $n$ correspond to the
lowest alternating parity bands, with $n=0$ being the yrast band, $n=1$
corresponding to the next non-yrast alternating parity structure and so on. The
values of the ``angular'' quantum number $k$ are not restricted and should only
satisfy the parity condition in i). The particular values of $k_n^{(+)}$ and
$k_n^{(-)}$ can be determined so as to reproduce the experimentally observed
parity shift in the set of levels with a given $n$.

iii) Due to the coherent interplay between the $\beta_2$ and $\beta_3$
variables in the oscillation motion, the excited $\beta$-bands in even-even
nuclei can be interpreted as the positive-parity counterparts of higher
negative-parity sequences, or as the members of non-yrast alternating-parity
bands.

Based on the above assumptions the extended alternating-parity spectrum of an
even-even nucleus can be considered in the following form.
\smallskip

\noindent Yrast alternating-parity set $(n=0)$: unites the $g$-band
$(k=k_{0}^{(+)})$ $I_{\nu}^{\pi}=0_{1}^{+},2_{1}^{+},4_{1}^{+}, 6_{1}^{+},...$
with the first negative-parity band denoted here as $n1$ $(k=k_{0}^{(-)})$
$I_{\nu}^{\pi}=1_{1}^{-},3_{1}^{-},5_{1}^{-},...;$
\smallskip

\noindent First non-yrast set $(n=1)$: unites the first $\beta$-band denoted by
$b1$  $(k=k_{1}^{(+)})$ $I_{\nu}^{\pi}=0_{2}^{+},2_{2}^{+},4_{2}^{+},...$ with
the second negative-parity band denoted by $n2$  $(k=k_{1}^{(-)})$
$I_{\nu}^{\pi}=1_{2}^{-},3_{2}^{-},5_{2}^{-},...;$
\smallskip

\noindent Second non-yrast set $(n=2)$: unites the second $\beta$-band $b2$
$(k=k_{2}^{(+)})$  $I_{\nu}^{\pi}=0_{3}^{+},2_{3}^{+},4_{3}^{+},...$ with the
third negative-parity band $n3$ $(k=k_{2}^{(-)})$
$I_{\nu}^{\pi}=1_{3}^{-},3_{3}^{-},5_{3}^{-},...$, and so on, higher non-yrast
sequences, where $\nu=1,2,3,...$ is the consequent number of the appearance of
a state with a given angular momentum. Also, it is convenient to use the band
labels introduced above to denote the different excited states as for example
$2_{g}^{+}$, $1_{n1}^{-}$, $0_{b1}^{+}$, $1_{n2}^{-}$ etc.

Obviously the above model scheme makes no claim to exhaust the entire
collective spectrum but rather provides a tool to identify the extent to which
the considered quadrupole-octupole motion can influence the excited band
structures in even-even nuclei. In the end of this section, it should be
remarked that the extension of the model to higher energy levels, together with
assumption ii), which releases $k$ from the fixed values $k^{(+)}=1$ and
$k^{(-)}=2$ (originally imposed in \cite{b2b3mod} for the yrast case), now
requires a new readjustment of the model parameters.

\section{Transition probabilities in the non-yrast quadrupole-octupole states}

As the B(E1) and B(E3) reduced transition probabilities are known to provide a
sensitive test for the structure of the alternating-parity sequences it is of
special importance to examine their behaviour in the non-yrast part of the
spectrum. The basic CQOM concept for the electromagnetic transitions has been
given in \cite{b2b3mod}. Here the formalism is modified so as to describe
B(E1), B(E2) and B(E3) reduced transition probabilities in the higher lying
alternating-parity bands along with the extended treatment of the model energy
quantum numbers. A more essential modification is related to a generalization
of the angular part of the electric transition operators dictated by the
complicated quadrupole-octupole shape density distribution inherent for the
coherent motion mode (see below). In addition the E1-E3 charge factors are
treated explicitly and the model parameters $p$ and $q$ (\ref{pqd}) providing
information about the potential shape are considered without including them
into scaling constants.

The reduced transition probability for an electric transition with a given
multipolarity $\lambda$ between model states (\ref{wftot}) with $n=n_i$,
$k=k_i$, $I=I_i$ and $n=n_f$, $k=k_f$, $I=I_f$ is
\begin{eqnarray}
B(E\lambda;n_i k_i I_{i}\rightarrow n_f k_f I_{f}) =
\frac{1}{2I_{i}+1}\sum_{M_{i}M_{f}\mu}\left| \left\langle
\Psi^{\pi_f}_{n_fk_fI_fM_f0}(\eta ,\phi)|\mathcal{M}_{\mu}(E\lambda)
|\Psi^{\pi_i}_{n_ik_iI_iM_i0}(\eta ,\phi) \right\rangle \right |^{2}.
\label{betrangen}
\end{eqnarray}
The operators for electric E1, E2 and E3 transitions have the following general
form
\begin{eqnarray}
\mathcal{M}_{\mu}(E\lambda)=\sqrt{\frac{2\lambda +1}
{4\pi (4-3\delta_{\lambda,1})}} \hat{Q}_{\lambda 0} D^{\lambda}_{0\mu},
\ \ \lambda=1,2,3,\ \ \mu=0,\pm 1, ..., \pm\lambda.
\end{eqnarray}
The vibration parts of these operators are given up to the first order of
$\beta_{2}$ and $\beta_{3}$, for E2 and E3, and in second order, for E1, as
\begin{eqnarray}
\hat{Q}_{1 0}&=&M_{1}\beta_{2}\beta_{3}
\label{q10} \\
\hat{Q}_{\lambda 0} &=& M_{\lambda}\beta_{\lambda}, \ \ \ \
\lambda = 2,3 \label{qL0}.
\end{eqnarray}
The electric charge factors $M_{\lambda}$  $(\lambda = 2,3)$ are taken as
\cite{LC88}
\begin{eqnarray}
M_{\lambda}=\frac{3}{\sqrt{(2\lambda +1)\pi}}ZeR_{0}^{\lambda},
\qquad \lambda = 2,3 \ ,
\label{multcrg}
\end{eqnarray}
where $R_{0}=r_0A^{1/3}$, $r_0\approx 1.2$ fm, $Z$ is the proton number, and
$e$ is the electric charge of the proton. The charge factor $M_{1}$ is taken
according to the droplet model concept \cite{MS74}-\cite{DMS86} in the form
\cite{DD95}
\begin{eqnarray}
M_{1}=\frac{9AZe^{3}}{56\sqrt{35}\pi}
\left( \frac{1}{J}+\frac{15}{8QA^{\frac{1}{3}}}\right ),
\label{m1d}
\end{eqnarray}
where the quantities $J$ and $Q$ are related to the volume and surface symmetry
energy, respectively and their values are assumed in the limits $25\lesssim
J\lesssim 44$ MeV and $17\lesssim Q\lesssim 70$ MeV \cite{BN91} (see also the
values below Eq.~(79) in \cite{DD95}). In the present work fixed average values
of these quantities $ J=35$ MeV and $Q=45$ MeV are used for all considered
nuclei. One should remark that so far there is no unique approach to estimate
the factor $M_{1}$. Therefore, here in (\ref{m1d}) the proton charge $e$ is
replaced by an effective charge $e^{1}_{eff}$ which is considered as an
adjustable parameter and can be different from one.  Note that to obtain the
B(E1) transition probabilities in the units $e^2\cdot \mbox{fm}^2$, and
subsequently in Weisskopf units one has to take into account that
$e^2=1.4399764\, \mbox{MeV}\cdot \mbox{fm}$ [or
$e^6/\mbox{MeV}^2=(1.4399764)^2e^2\cdot \mbox{fm}^2$] which leads to an
additional multiplication factor 1.4399764 in (\ref{m1d}) when numerical values
are produced.

In the space of the ellipsoidal coordinates (\ref{polar}), (\ref{pqd}) one has
\begin{eqnarray}
\hat{Q}_{1 0}&=&M_{1}pq\eta^{2}\cos\phi\sin\phi \label{q1cs}\\
\hat{Q}_{2 0}&=&M_{2}p\eta\cos\phi \label{q2c}\\
\hat{Q}_{3 0}&=&M_{3}q\eta\sin\phi \label{q3s}.
\end{eqnarray}
The definitions of the operators (\ref{q10})--(\ref{qL0}) and
(\ref{q1cs})--(\ref{q3s}) originally correspond to a situation in which the
nuclear shape is characterized by fixed values of the deformation parameters
$\beta_{2}$ and $\beta_{3}$. In this case the density distribution of the
collective state is characterized by a single maximum in the space of
$\beta_{2}$ and $\beta_{3}$. In the case of the model potential (\ref{Ub2b3I})
taken with an elliptic bottom the density distribution can be characterized by
more than one maximum. Indeed, the density of the model state (\ref{wvib}) is
characterized by a different number of maxima depending on the quantum number
$k$. This feature is a result of the assumed soft quadrupole--octupole mode. It
is illustrated graphically in Appendix A, where the density distribution of the
state (\ref{wvib}) in the space of the quadrupole-octupole deformations is
plotted for different $k$-values at $n=0$ after transforming the wave function
$\Phi^{\pi}_{nkI}$ in the ($\beta_{2}$,$\beta_{3}$) variables. It is seen that
for $\beta_{2}>0$ the number of maxima is equal to $k$ and by analogy with the
acoustics may be interpreted as the number of ``overtones'' which characterize
the coherent collective oscillations of the system. Thus, it appears that the
transition operators should connect states with different numbers of maxima (or
overtones). In the space of ellipsoidal variables the positions of the maxima
are determined by the angular variable $\phi$. On the other hand the original
operators (\ref{q1cs})--(\ref{q3s}) do not take into account the presence of
multiple maxima in the shape density distributions of the different states. One
particular effect due to this circumstance is that the integrals over the
angular part of (\ref{q3s}), $\sin\phi$, vanish when the difference between the
$k$ numbers of the initial and final states is larger than a unit and the
respective B(E3) transition probabilities vanish too. This limitation is
removed if the operators are generalized appropriately. The most general forms
of the angular parts of the operators corresponding to the first orders of
$\beta_{2}$ and $\beta_{3}$ according to (\ref{q2c}) and (\ref{q3s}) can be
sought in terms of a Fourier expansion with respect to $\phi$ through the
replacements
\begin{eqnarray}
\cos\phi\rightarrow A_{2 0}(\phi)\equiv \sum_{k=1}^\infty a^{(k)}_{2 0}
\cos (k\phi),\qquad   \sin\phi\rightarrow A_{3 0}(\phi)\equiv
\sum_{k=1}^\infty a^{(k)}_{3 0} \sin (k\phi ),
\end{eqnarray}
where the expansion coefficients should be chosen so as to provide an
appropriate convergence. A choice made here for both type of coefficients is
$a^{(k)}=1/k$ for which the above expansions can be obtained in analytic form
\begin{eqnarray}
A_{2 0}(\phi)&=&\sum_{k=1}^\infty \frac {\cos (k\phi )}{k}
=-\frac{1}{2}[\ln 2 +\ln(1-\cos \phi)] \label{a20}\\
A_{3 0}(\phi)&=&\sum_{k=1}^\infty \frac {\sin (k\phi )}{k}
=\frac{\pi- \phi}{2} + \pi \mbox{Floor}\left(\frac{\phi}{2\pi}\right),
\label{a30}
\end{eqnarray}
where the Floor function maps a real number to the largest previous integer.
Then the angular part of the second order operator (\ref{q1cs}) can be
generalized in an obvious way
\begin{eqnarray}
\cos\phi\sin\phi\rightarrow A_{1 0}(\phi)\equiv
A_{2 0}(\phi)A_{3 0}(\phi)=
\sum_{m=1}^\infty\sum_{n=1}^\infty \frac {\cos (m\phi )}{m}
\frac{\sin (n\phi )}{n}.
\label{a10}
\end{eqnarray}
Note that the first terms of the above expansions represent the original
angular parts in (\ref{q1cs})--(\ref{q3s}). So, the new angular operators
(\ref{a20})--(\ref{a10}), which are extensions of the old ones, provide a
connection between states whose ``dynamical'' deformations (i.e. the
probability distribution in the deformation space) are characterized by the
co-existence of a large number of maxima. These specific shape properties of
the system are due to the assumed coupling between quadrupole and octupole
degrees of freedom. Now the operators (\ref{q1cs})--(\ref{q3s}) are redefined
as
\begin{eqnarray}
\hat{Q}_{1 0}(\eta ,\phi)&=& M_{1}pq\eta^{2} A_{1 0}(\phi) \label{q1gen}\\
\hat{Q}_{2 0}(\eta ,\phi)&=& M_{2}p\eta A_{2 0}(\phi)\label{q2gen}\\
\hat{Q}_{3 0}(\eta ,\phi)&=& M_{3}q\eta A_{3 0}(\phi)\label{q3gen}.
\end{eqnarray}

After carrying out the integration over the rotation part in (\ref{betrangen})
one obtains
\begin{eqnarray}
B(E\lambda;n_i k_i I_{i}\rightarrow n_f k_f I_{f}) =
\frac{2\lambda +1}{4\pi (4-3\delta_{\lambda,1})}
\langle I_i0\lambda 0|I_f0\rangle^2
R_{\lambda}^{2}(n_i k_i I_{i}\rightarrow n_f k_f I_{f}),
\end{eqnarray}
which involves the squares of the Clebsch-Gordan coefficient and the matrix
element of the electric multipole operators (\ref{q1gen})-(\ref{q3gen}) between
the quadrupole-octupole vibration wave functions (\ref{wvib})
\begin{eqnarray}
R_{\lambda}(n_i k_i I_{i}\rightarrow n_f k_f I_{f})
=\left\langle \Phi^{\pi_f}_{n_f k_f I_f} (\eta,\phi) |\hat{Q}_{\lambda 0} |
\Phi^{\pi_i}_{n_i k_i I_i}(\eta,\phi) \right\rangle.
\label{rlam}
\end{eqnarray}
By further separating the integrations over the ``radial'' variable  $\eta$ and
the ``angular'' variable  $\phi$ in (\ref{rlam}) according to (\ref{wvib}) one
obtains
\begin{eqnarray}
R_{1}(n_i k_i I_{i}\rightarrow n_f k_f I_{f})
&=&M_{1}pqS_{2}(n_i,I_i;n_f,I_f)I_{1}^{\pi_i,\pi_f}(k_i,k_f) \label{rlam1}\\
R_{2}(n_i k_i I_{i}\rightarrow n_f k_f I_{f})
&=& M_{2}pS_{1}(n_i,I_i;n_f,I_f)I_{2}^{\pi_i,\pi_f}(k_i,k_f) \label{rlam2}\\
R_{3}(n_i k_i I_{i}\rightarrow n_f k_f I_{f})
&=& M_{3}qS_{1}(n_i,I_i;n_f,I_f)I_{3}^{\pi_i,\pi_f}(k_i,k_f),
\label{rlam3}
\end{eqnarray}
where
\begin{eqnarray}
S_{1}(n_i,I_i;n_f,I_f)&=&\int_0^{\infty} d\eta
\psi_{n_f}^{I_f}(\eta)\eta^{2} \psi_{n_i}^{I_i}(\eta)
\label{s1} \\
S_{2}(n_i,I_i;n_f,I_f)&=&\int_0^{\infty} d\eta
\psi_{n_f}^{I_f}(\eta)\eta^3 \psi_{n_i}^{I_i}(\eta),
\label{s2}
\end{eqnarray}
and
\begin{eqnarray}
I_{\lambda}^{\pi_i,\pi_f}(k_i,k_f)&=&\frac{2}{\pi}\int_{-\frac{\pi}{2}}^{\frac{\pi}{2}}
A_{\lambda 0}(\phi)\varphi^{\pi_f}_{k_f}(\phi)\varphi^{\pi_i}_{k_i}(\phi)d\phi ,
\ \ \ \lambda =1,2,3.
\label{Ilam}
\end{eqnarray}

The integrals over $\eta$,  (\ref{s1}) and (\ref{s2}), involve the ``radial''
wave functions (\ref{psieta1}). Analytic expressions for these integrals are
given in Appendix B. The integrals over $\phi$ (\ref{Ilam}) involve the
``angular'' wave functions (\ref{parplus}) and/or (\ref{parminus}). The
explicit forms of these integrals with the relevant parities $\pi_{i}$ and
$\pi_{f}$ are given in Appendix C.

From the generalized definitions (\ref{q1gen})-(\ref{q3gen}) of the operators
$\hat{Q}_{\lambda 0}$ it is seen that the inertial factors $p$, $q$ and their
product $pq$ defined through Eq.~(\ref{pqd}) are not included in any scaling
factors, as done in Ref.~\cite{b2b3mod} and can be considered as model
parameters. Actually, $p$ and $q$ are not independent. From (\ref{pqd}) it can
be easily seen that $1/p^{2}+1/q^{2}=2$. Then $q$ can be expressed by $p$ as
\begin{eqnarray}
q=\frac{p}{\sqrt{2p^{2}-1}}, \qquad \mathrm{with}
\qquad  p>\frac{1}{\sqrt{2}}\approx 0.7071.
\label{q}
\end{eqnarray}
The inequality in (\ref{q}) corresponds to the condition $d_2<2d$. Analogically
one can express $p$ by $q$ with the condition $d_3<2d$. (Note that for $p=1 $
one has $q=1$ and $pq=1$.) Here the adjustable parameter is chosen to be $p$.
It should be noted that with the involvement of the new parameter $p$ the
scaling factors in Eqs.~(46) and (47) of Ref.~\cite{b2b3mod} are not considered
anymore and the charge factors $M_2$ and $M_3$ are directly calculated in
(\ref{multcrg}). Also the charge factor $M_1$ is directly calculated in
(\ref{m1d}), but with the effective charge $e^{1}_{eff}$ being adjusted to
determine the correct scale of the B(E1) transition probabilities with respect
to B(E2). From another side the parameter $p$ determines the relative scale
between B(E2) and B(E3). It is interesting to remark that $p$ does not play any
role if the model energy levels are fitted without taking into account
transition probabilities. However, in this case there is an ambiguity in the
choice of the inertial parameters $d_2$ and $d_3$. This is seen by the
following relations between the parameters of the potential (\ref{Ub2b3I}) and
the fitting parameters $\omega$ and $b$, imposed by the coherent condition
(\ref{om})
\begin{eqnarray}
C_2&=& \frac{\omega^{2}b}{2}d_2, \qquad  C_3= \frac{\omega^{2}b}{2}d_3
\label{C2C3}\\
B_2&=& \frac{b}{2}d_2, \qquad\ \ \ B_3= \frac{b}{2}d_3,
\label{B2B3C2C3}
\end{eqnarray}
in which $d_2$ and $d_3$ are not determined. (The parameter $d_0$ does not
enter these relations.) This means that for a given set of parameters $\omega$,
$b$ and $d_0$ the energy spectrum corresponds to an infinite number of
potential shapes with different eccentricities of the ellipsoidal bottom. Now,
after determining the parameters $p$ and $c$ with respect to transition data
one gets
\begin{eqnarray}
d_2=\frac{d}{p^{2}}, \qquad d_3=(2p^{2}-1)\frac{d}{p^{2}},
\qquad \mathrm{with}  \qquad d=\frac{2c}{\omega b}.
\label{d2d3}
\end{eqnarray}
Thus for given values of the parameters $\omega$, $b$, $c$ and $p$ the original
parameters $B_2$, $B_3$, $C_2$, $C_3$, $d_2$ and $d_3$ of potential
(\ref{Ub2b3I}) are fixed, and given additionally $d_0$, its form is
unambiguously determined.

\section{Numerical results and discussion}

The extended CQOM formalism was applied to several nuclei, namely
$^{152,154}$Sm, $^{154,156,158}$Gd, $^{236}$U and $^{100}$Mo, in which one or
two non-yrast alternating-parity bands can be constructed by the experimentally
observed $\beta$ and higher-lying negative-parity levels. In these nuclei a
number of data on E1 and/or E3 transitions are available providing the
possibility to test the complete model scheme. In all selected nuclei the
experimental data \cite{ensdf} provide well determined yrast and first
non-yrast alternating-parity bands except for $^{100}$Mo where the structure of
the non-yrast band is proposed here on the basis of the model analysis (see
below). In three of the nuclei, $^{154}$Sm, $^{154}$Gd and $^{158}$Gd second
excited (non-yrast) alternating-parity bands are additionally considered. The
structure of these bands is not clearly determined in the experimental data.
Therefore, the model description and prediction provides a possible
interpretation of the respective experimental levels. In this meaning the
present description not only provides a test for the CQOM model scheme, but
also suggests a possible classification of some highly non-yrast excited states
whose interpretation in the experimental data bases is not unambiguous.

The model description is obtained by taking the theoretical energy levels
$\tilde{E}_{n,k}(I) =E_{n,k}(I)-E_{0,k_{0}^{(+)}}(0)$ from Eq.~(\ref{enspect}).
The parameters $\omega$, $b$, $d_{0}$, $c$, $p$ and the effective charge
$e^{1}_{eff}$ have been adjusted by simultaneously taking into account
experimental data on the energy bands and the available B(E1)-B(E3) transition
probabilities. The parameter values obtained in the considered nuclei are given
in Table 1. The resulting values of the original Hamiltonian parameters in
(\ref{Hqo}) and (\ref{Ub2b3I}) are given in Table~2. For each nucleus the
calculations are performed in a net over the values of  the ``angular'' quantum
numbers $k$ with appropriate parity in the limits $1\leq k\leq 20$. In all
nuclei sets of values for the $k$-quantum numbers providing the best model
description of both, energies and reduced transition probabilities, are
obtained. These values are given in Figures 1--7 where the theoretical and
experimental energy levels of the considered nuclei are compared. The
theoretical and experimental values of the B(E1), B(E2) and B(E3) transition
probabilities are compared in Table 3. Model predictions for some not yet
observed transitions are also given there.

The results in Figs. 1--7 show that the model scheme correctly reproduces the
structure of the alternating-parity spectra in the considered nuclei with a
reasonably good agreement between the theoretical and experimental energy
levels. The correct reproduction of the mutual displacement of the different
positive and negative parity sequence is related to the involvement of $k$
quantum number values larger than 1 and 2. On the other hand the determination
of the $k$-values is strongly dictated by the interband transitions between the
positive- and negative- parity levels as well as by the transitions between the
different alternating-parity sequences.  The above remarks explain why for the
nuclei $^{152}$Sm and $^{154}$Gd new sets of $k$-quantum numbers appear
together with renormalized values of the fitting parameters compared to the
previous descriptions limited to the yrast bands \cite{b2b3mod} (see below).
One should remark that at the same time the main (radial) oscillator quantum
number $n$ is uniquely determined, $n=0$ for the yrast sequence, $n=1$ for the
first excited alternating parity band and so on, as explained in the end of
Sec. 2.

In $^{152}$Sm the yrast band is described together with the first excited band
(see Fig.~1). The calculations provide two identical couples of $k$ values
$(k^{(+)}=1,k^{(-)}=8)$ for each band. Thus it is seen that $k^{(-)}$ obtains a
value larger than the lowest even value $2$ considered in \cite{b2b3mod}. From
Table~3 one can see that with this configuration of $k$-numbers the model
fairly good reproduces the data \cite{nudat2} on the B(E2) intraband transition
probabilities in the ground-state band ($g$) and on the B(E1) probabilities for
transitions between the $g$- and the first negative-parity band ($n1$).  Some
interband $E2$ transitions between the $g$- and the first $\beta$-band ($b1$),
as $2_{b1}^{+}\rightarrow 2_{g}^{+}$ and $4_{b1}^{+}\rightarrow 4_{g}^{+}$ are
also well described, while others like $4_{b1}^{+}\rightarrow 6_{g}^{+}$ are
overestimated. The calculated intraband transitions in the $b1$-band are in
rough agreement with the experimental data, while the $E1$ intraband transition
$1_{n1}^{-}\rightarrow 2_{b1}^{+}$ is overestimated by an order. The E3
transition probability B$(E3;3_{n1}^{-}\rightarrow  0_{g}^{+})=14$ W.u.
\cite{K02} is exactly reproduced due to the adjustable parameter $p$ which
determines the factor $q$ in (\ref{rlam3}) according to Eq.~(\ref{q}). This
allows one to predict other E3 transitions like $1_{n1}\rightarrow 4_{g}$ and
other similar transitions between the $b1$-band and the second negative-parity
band ($n2$) as shown in Table~3. Although not all theoretical transition
probabilities are in strict agreement with the experimental data it is seen
that the model scheme correctly takes into account the different scales of the
various kinds of probabilities. A similar behaviour of transition probabilities
is observed in the other considered nuclei.

In $^{154}$Sm totally three alternating-parity bands are considered as seen
from Fig.~2. The positive-parity states of the second excited band are
interpreted in \cite{ensdf} as members of a second $K^{\pi}=0^+$ band, or of a
second $\beta$-band $(b2)$. The respective negative-parity levels are selected
in the present work among levels for which there is no interpretation given in
\cite{ensdf}. Here they form a third negative-parity band $(n3)$. From Table~3
it is seen that the intraband B(E2) transition probabilities in the $g$-band of
this nucleus are reasonably well described up to $I=10$, while the
B$(E2;12_{g}^{+}\rightarrow  10_{g}^{+})$ value is considerably overestimated.
The B(E1) probabilities between the $n1$- and $g$-bands are also well
described. The theoretical interband transition value
B$(E2;0_{b1}^{+}\rightarrow  2_{g}^{+})$ is by an order smaller than the
experimental one. For the other similar E2 transitions like
$2_{b1}^{+}\rightarrow  0_{g}^{+}$ the theoretical values are obtained below
the upper limits given for the respective experimental data \cite{nudat2}.

In $^{154}$Gd, again, three alternating-parity bands are considered. The model
description is given in Fig.~3. Here, the second non-yrast band is constructed
by the second excited $K^{\pi}=0^+$ band and a $3^{-}$ state with energy
1796.96 keV \cite{ensdf}. Although the latter is interpreted in \cite{ensdf} as
a member of a $K^{\pi}=2^-$ octupole band it reasonably fits the present scheme
as a member of an $n3$-sequence. In this nucleus the B(E1)--B(E3)  transition
probabilities are also reasonably described with the largest discrepancies
between the theory and experiment, about a factor of 2, being observed for the
E2 transitions $2_{b1}^{+}\rightarrow  0_{b1}^{+}$ and $0_{b1}^{+}\rightarrow
2_{g}^{+}$ (see Table~3). Note that here the theoretical B(E1) value for the
interband transition B$(E1;1_{n1}^{-}\rightarrow 2_{b1}^{+})=0.0064$ W.u. is
obtained close to the experimental one, $0.0099$ W.u.

In $^{156}$Gd two alternating-parity bands, the yrast and first excited, are
considered (see Fig.~4). The B(E2) and B(E1) transition probabilities between
the members of the $g$-, $b1$- and $n1$-bands are well described with a few
exceptions as in the transitions $4_{b1}^{+}\rightarrow  2_{b1}^{+}$ and
$4_{b1}^{+}\rightarrow  2_{g}^{+}$ for which the B(E2)-values are
underestimated with respect to the experiment by a factor of about two and an
order, respectively (see Table 3). On the other hand the model predictions for
the B(E1) transition probabilities between the second negative-parity band $n2$
and the $g$-band suggest $2$-$3$ orders of magnitude in suppression compared to
experimental data.

In $^{158}$Gd, three alternating-parity bands are considered (see Fig.~5).
Similarly to $^{154}$Gd the $1^{-}$ and $3^{-}$ states included in the second
excited band enter the present model scheme as $n3$-members, while in
\cite{ensdf} they are interpreted as members of a $K^{\pi}=1^-$ octupole band.
For this nucleus, quite a large number of data on B(E1) and B(E2) transition
probabilities are available \cite{nudat2}. One should remark that compared to
the other considered rare-earth nuclei $^{158}$Gd is closer situated to the
region of pronounced rotation collectivity. From Table~3 it is seen that the
theoretical intraband B(E2) probabilities in the $g$-band of $^{158}$Gd faster
increase with the angular momentum compared to the experimental data. On the
other hand six experimental B(E1) values for the transitions between the $g$-
and the $n1$-bands are described quite well. It is remarkable that an
experimental estimation for a E1 transition between the $n2$- and $b1$-band is
available with B$(E1;3_{n2}^{-}\rightarrow  2_{b1}^{+})>0.00035$ W.u. This
circumstance is in a conjunction with the model assumption about the
quadrupole-octupole coupling of both bands. The model description predicts for
this probability a smaller value of 0.00011 W.u., which is of the same order as
the B(E1) values connecting the $g$- and $n1$-bands. Further, model prediction
values for similar B(E1) transition probabilities as
B$(E1;1_{n2}^{-}\rightarrow 0_{b1}^{+})=8\times 10^{-5}$ W.u. and
B$(E1;1_{n2}^{-}\rightarrow 2_{b1}^{+})=0.0002$ W.u. are given in Table~3. Also
there one can find available experimental estimations for intraband transition
probabilities like B$(E2;5_{n1}^{-}\rightarrow 3_{n1}^{-})=369$ W.u. and
B$(E2;3_{n2}^{-}\rightarrow 1_{n2}^{-})>1600$ W.u. which are underestimated by
the theory. In addition, a number of B(E1) transition probabilities from $b1$-
to $n1$-, from $n2$- to $g$- and from $b2$- to $n1$- and $n2$-bands are
generally underestimated by one or two orders of magnitude.

In $^{236}$U two alternating-parity bands, the yrast and first excited, are
considered (see Fig.~6). This nucleus was selected because of the possibility
to examine two observed reduced probabilities for E3 transitions, namely
B$(E3;1_{n1}^{-}\rightarrow  4_{g})=62$ W.u. \cite{nudat2} and
B$(E3;3_{n1}\rightarrow  0_{g})=22.9$ W.u. \cite{K02}. From Table~3 it is seen
that the first one is exactly reproduced. The second one is underestimated by
the theoretical value, $15$ W.u., which is still reasonably close to the
experiment. For the B(E2) transition probabilities within the $g$-band the
description is good as overall up to a quite high angular momentum $I=26$. The
experimental B(E1) value for the transition $1_{n1}^{-}\rightarrow  0_{g}^{+}$
is exactly reproduced because of the use of the effective charge. This allows
one to predict other B(E1) transition probabilities in the described spectrum
which are given in Table~3.

In $^{100}$Mo the experimentally observed $2_{3}^{+}$ state with energy
$1463.9$ keV is considered in \cite{ensdf} as a possible member of a
$\beta$-band. However, the present scheme suggests that the $2^{+}$ state
belonging to this band should lie essentially lower. The calculations show that
the experimental $2_{2}^{+}$ state with energy $1063.78$ keV considered in
\cite{ensdf} as a possible member of a $\gamma$-band is more appropriate as a
$\beta$-band member. The result in Fig.~7 shows that if this state is included
in the $b1$-band (in the present notations) a non-yrast alternating-parity
sequence can be constructed and reasonably well described by taking three
additional states, namely $1^-$ at $2156$ keV, $3^-$ at $2369.6$ keV and $4^+$
at $1771.4$ keV from the set of available but not interpreted data for
$^{100}$Mo \cite{ensdf}. The observed B(E1)-B(E3) transition probabilities are
reasonably described as seen from Table~3. The main discrepancy between the
theory and the experiment, a factor of 5, is obtained for the E2 intraband
transition $2_{b1}^{+}\rightarrow  0_{b1}^{+}$.

The following comments on the model results can be made here. The parameters of
the fits shown in Table~1 reflect the common collective structure of the
various energy sequences ($g$, $b1$, $b2$, $n1$, $n2$, $n3$) in a given
nucleus, while the sets of $k$ values given in Figs 1-7 reflect their mutual
dispositions. Note that the parameters for $^{152}$Sm and $^{154}$Gd are
essentially renormalized compared to the fits of the yrast band only
\cite{b2b3mod}. As seen from Table~1 the parameters $\omega$ and $b$, which are
responsible for the rotation-vibration behaviour of the different sequences,
vary relatively smoothly between the different nuclei. The parameter $d_0$,
which is responsible for the shape of the potential at zero angular momentum,
shows more pronounced differences in its values, especially for the nuclei from
different regions as $^{236}$U and $^{100}$Mo. Also, the parameter $c$, which
determines the overall scale for the transition probabilities in the ``radial''
integrals, considerably varies, while the parameter $p$ which is related to the
quadrupole and octupole contributions to the moment of inertia changes quite
smoothly. It is remarkable that in three nuclei, $^{152}$Sm, $^{154}$Sm and
$^{154}$Gd, the effective charge for the E1 transitions is practically unit
which means that there is no need for this parameter to describe them. In
$^{156}$Gd it is still close to 1, while in the other three nuclei its need for
the model description is already essential.

By using the relations (\ref{B2B3C2C3}) and (\ref{d2d3}) between the model
parameters in ellipsoidal coordinates and the parameters of the original
Hamiltonian, (\ref{Hqo}) with (\ref{Ub2b3I}), one can obtain the latters from
the values given in Table~1. Subsequently one can obtain the semi-axes (sa)
$\beta_{2}^{\mbox{\scriptsize sa}}$ and $\beta_{3}^{\mbox{\scriptsize sa}}$ of
the ellipsoidal potential bottom in the space of the quadrupole--octupole
variables given by
\begin{eqnarray}
\beta_{\lambda}^{\mbox{\scriptsize sa}}(I)=[2X(I)/d_{\lambda}C_{\lambda}]^{1/4},
\qquad \lambda=2,3.
\label{semiax}
\end{eqnarray}
(For more details see the text after Eqs. (3) and (4) of \cite{b2b3mod}.) The
resulting values of the parameters $B_2$, $B_3$, $C_2$, $C_3$, $d_2$, $d_3$ and
the semi-axes are given in Table~2. Note that in the present work they are {\em
not} directly adjusted, but obtained as a result of the adjustment of the
parameters $\omega$, $b$, $d_0$,  $c$, $p$, and $e^{1}_{eff}$. As such they
only give a rough estimation about the order of the potential parameters and
its shape. One can see that for $^{152,154}$Sm and $^{154,156}$Gd these
parameters vary relatively smoothly, while for the remaining three nuclei they
show some essential fluctuations. The values of the
$\beta_{2}^{\mbox{\scriptsize sa}}$ semi-axis are obtained close to the known
values of the static quadrupole deformations in these nuclei while the values
of the octupole semi-axis $\beta_{3}^{\mbox{\scriptsize sa}}$ appear
considerably larger. This result is correlated with the larger values of the
quadrupole stiffness parameters $C_2$ compared to the values of $C_3$. Hence
the present parameters correspond to a vibration motion with a larger softness
of the system with respect to the octupole mode compared to the quadrupole one.
A closer look on the formalism shows that the ratio between both semi-axes is
related to the matrix elements of the quadrupole and octupole electric
multipole operators (\ref{q2gen}) and (\ref{q3gen}). By using (\ref{C2C3}),
(\ref{d2d3}) and (\ref{q}) in (\ref{semiax}) one finds that
\begin{eqnarray}
\frac{\beta_{3}^{\mbox{\scriptsize sa}}}{\beta_{2}^{\mbox{\scriptsize sa}}}
=\frac{q}{p}=\frac{1}{\sqrt{2p^{2}-1}}.
\label{semrat}
\end{eqnarray}
It is seen that the ratio $\beta_{3}^{\mbox{\scriptsize
sa}}/\beta_{2}^{\mbox{\scriptsize sa}}$ depends on the inertia factors $p$ and
$q$, Eq~(\ref{pqd}), which determine the strength of the E2 and E3 transitions,
respectively. This ratio is less than 1 for $p>1$ $(q<1)$. It can be easily
checked that to obtain $\beta_{3}^{\mbox{\scriptsize
sa}}/\beta_{2}^{\mbox{\scriptsize sa}}<1$ one has to introduce an additional
scaling constant, $c_3$, having the meaning of an effective charge for the
octupole mode. Then the octupole charge factor is renormalized as
$M_{3}'=c_3M_{3}$. The numerical analysis shows that if $c_3$ is chosen in the
limits $2\leq c_3\leq4$ the parameter $p$ is renormalized so that $q\rightarrow
q/3$ and the same theoretical levels and transition probabilities are obtained
with $\beta_{3}^{\mbox{\scriptsize sa}}<\beta_{2}^{\mbox{\scriptsize sa}}$ in
correspondence to the usually observed values of the deformation parameters
$\beta_{2}$ and $\beta_{3}$. For example if $c_3=4$ one obtains the following
set of renormalized parameters for $^{154}$Gd, $c'=269.6$, $p'=1.197$,
${e^{1}_{eff}}'=1.512$, while the parameters $\omega$, $b$ and $d_{0}$ remain
unchanged compared to the values given in Table~1. Compared to the values in
Table~2 the renormalized parameters for $^{154}$Gd are ${B_3}'=1146$
$\hbar^{2}/$MeV, ${C_3}'=108$ MeV, ${d_3}'=777$ $\hbar^{2}/$MeV and
${\beta_{3}^{\mbox{\scriptsize sa}}}'=0.192$, while the other parameters
referring to the quadrupole deformation remain unchanged. It is seen that now
the length of the potential bottom semi-axis in the $\beta_{3}$-direction
corresponds to a more realistic octupole deformation. This result is equivalent
to the involvement of a renormalized octupole operator $\hat{Q}_{3 0}'(\eta
,\phi)=c_3\hat{Q}_{3 0}(\eta ,\phi)$. Since the use of such an effective charge
does not change the model description but only leads to the renormalization of
the parameters it is not considered in the present work.

Further, it is important to comment the obtained configurations of quantum
numbers $k_n^{(+)}$ and  $k_n^{(-)}$ which characterize the energy shifts in
the described alternating-parity spectra. From Figs.~1--7 it is seen that the
relevant energy shift in the excited level sequences is obtained through a jump
of $k$ over several lower values. In this way certain low-lying states
available in the scheme do not enter the considered spectrum, while others
lying at higher energy are used to obtain the model description. This result is
a consequence of the fact that the same oscillation frequency $\omega$ is
imposed to all alternating-parity bands. Actually, the non-yrast bandheads and
the energy shifts could be reproduced through the lowest possible
$k$-configurations [$k_n^{(+)}=1,k_n^{(-)}=2$] if separate vibration
frequencies are considered in the different bands. Speaking about $k$ as a
number of angular oscillation quanta (phonons) it appears that the restricted
freedom of the frequency imposed by the coherent condition is compensated in
the model description by the presence of a larger number of quanta on which the
rotation bands are built. Since the eventual consideration of different
oscillation frequencies would correspond to the introduction of parameters
external for the model the larger numbers of quanta are retained in the present
work. The obtained pairs of values $k_n^{(+)}$ and  $k_n^{(-)}$ for the quantum
number $k$ provide a detailed systematic information about the mutual
disposition of the positive- and negative-parity bands in the different nuclei,
and subsequently, about the evolution of the quadrupole-octupole spectra in a
given nuclear region. It should be noted that the involvement of the extended
transition operators (\ref{q1gen})-(\ref{q3gen}) in the present CQOM
development is related to the appearance of larger $k$-values and the
subsequent large $k$-differences taken into account in the electric transition
probabilities. These features of the model can change if it is applied beyond
the coherent-mode assumption. In this case the unrestricted Hamiltonian
(\ref{Hqo}) can be diagonalized by using the present analytic solution as a
basis. Then the parameters in (\ref{Hqo}) can be directly adjusted to describe
the spectrum without restriction of the quadrupole and octupole oscillator
frequencies. This could allow one to construct the spectrum by always choosing
the lowest possible eigenvalues, while the structure of the spectrum obtained
in the present analytic solution could only guide the construction of non-yrast
bands. Work in this direction is in progress.

Finally, it should be noted that the present model descriptions are obtained
within some natural limits of the applied formalism with respect to
experimental data. It is well known that rotation terms like the one entering
the model potential can only describe smooth changes of the rotation spectra
with increasing angular momentum, as for example the so called ``centrifugal
stretching''. The treatment of angular momentum regions where sharper changes
in the rotation spectrum due to changes in the intrinsic structure like
backbending effects occur, needs a special development which is not the subject
of the present work. That is why in some of the considered nuclei descriptions
and/or predictions of rotation levels with very high angular momenta are
avoided, especially in the cases where the negative-parity levels are not
observed. An exception is done for $^{236}$U (Fig.~6), where higher-spin
negative-parity levels were predicted in accordance to the last observed state
with even angular momentum. This prediction should be meaningful since in the
actinide region the rotation spectra exhibit more regular rotation motion in
the high-spin regions. On the other hand the prediction of missing low-spin
states, like the $1^{-}_{n3}$ level in $^{154}$Gd and the $6^{+}_{b1}$ and
$5^{-}_{n2}$ levels in $^{100}$Mo, as well as, a number of not observed
transition probabilities shown in Table~3 should be also reasonable in the
present framework. In this meaning the applied CQOM model scheme rather
describes the ``horizontal'' evolution of the alternating-parity spectra beyond
the yrast line than the high-spin properties of  individual rotation bands.

\section{Concluding remarks}

The present work provides a model description and respective classification of
the yrast and non-yrast alternating-parity spectra and the attendant B(E1),
B(E2) and B(E3) transition probabilities in several rare-earth nuclei, one U
and one Mo nucleus within the collective model of Coherent Quadrupole and
Octupole Motion (CQOM). The theoretical formalism and the obtained model
descriptions outline a possible way for the development of nuclear
alternating-parity spectra toward the highly non-yrast region of collective
excitations. In the considered scheme the different negative parity
level-sequences appear in couples together with the ground-state band and the
excited $\beta$-bands. On this basis the model predicts possible E(1) and E(3)
transitions between states with opposite parity within various
alternating-parity bands. The presence of experimentally observed E(1)
transitions between such states in the non-yrast part of the spectrum is
noticed. Further experimental measurements of electric transition probabilities
would be very useful to check the possible coupling of non-yrast energy
sequences with opposite parities. It was demonstrated that the considered
scheme can be used for the interpretation of data on excitation energies whose
place in the structure of the collective spectrum has not yet been determined.
The approach was applied to selected nuclei for which a relatively large number
of data on B(E1)-B(E3) transitional probabilities are available, but it can be
easily extended to wider ranges of nuclei especially in the rare-earth and
actinide regions. Further, the formalism takes into account the complex-shape
effects in the motion of the system and in addition provides estimations about
the shape of the quadrupole--octupole potential which governs the collective
properties of the considered nuclei. More refined model descriptions and
realistic estimations about the potential shape can be obtained beyond the
limits of the present coherent-mode assumption. Work in this direction is in
progress.

\section*{Acknowledgements}

\noindent We thank Professor Jerzy Dudek for valuable discussions and comments.
This work is supported by DFG and by the Bulgarian National Science Fund
(contract DID-02/16-17.12.2009).

\section*{\bf Appendix A: CQOM shape-density distributions}

The density distribution of the CQOM vibration state in the space of the
quadrupole-octupole shapes is given by the square of the wave function
(\ref{wvib}), $\rho_{nkI}(\beta_{2},\beta_{3})=|\Phi^{\pi}_{nkI}
(\beta_{2},\beta_{3})|^{2}$, after a transformation from the ellipsoidal
coordinates $(\eta ,\phi)$ to the deformation coordinates $(\beta_{2}
,\beta_{3})$. In Fig.~8 three-dimensional plots of $\rho_{nkI}$ are given for
the lowest $k=1$ and $k=2$ states for $n=0$ and for the schematic parameter
values $\omega =0.3\,$MeV/$\hbar$, $b=3\,\hbar^{-2}$, $d_0=100\,\hbar^{2}$,
$d_2=300\,\hbar^{2}$/MeV, $d_3=500\,\hbar^{2}$/MeV. Note that according to the
discussion in the end of Sec. 3 the shape of the potential is determined
unambiguously when the values of the inertia parameters $d_2$ and $d_3$ are
given. In Fig.~9 two-dimensional plots showing the maxima of $\rho_{nkI}$ for
$k=1-4$ are given together with contours showing the ellipsoidal potential
bottom for the above set of schematic parameters.

\section*{\bf Appendix B: Explicit form of the integrals over $\eta$}

The integrals over $\eta$, (\ref{s1}) and (\ref{s2}), can be written in the
following common form after taking into account the explicit expression for the
radial wave functions (\ref{psieta1})
\begin{eqnarray}
S_{l}(n_i,I_i;n_f,I_f)&=&\int_{0}^{\infty}d \eta
\psi_{n_{f}}^{I_{f}}(\eta)\eta^{l+1}\psi_{n_{i}}^{I_{i}}(\eta)
\nonumber \\
&=&N \int_{0}^{\infty}
e^{-c\eta^{2}}c^{s_{f}}\eta^{2s_{f}}L_{n_{f}}^{2s_{f}}(c\eta^{2})\eta^{l+1}
c^{s_{i}}\eta^{2s_{i}}L_{n_{i}}^{2s_{i}}(c\eta^{2})d \eta,
\label{Sleta}
\end{eqnarray}
where $l=1,2$, $s_{i}=(1/2)\sqrt{k_{i}^{2}+bX(I_i)}$,\ \
$s_{f}=(1/2)\sqrt{ k_{f}^{2}+bX(I_f)}$  and
\begin{eqnarray}
N=N_{n_i,n_f}(c,s_i,s_f)=2c\left[\frac{\Gamma(n_{f}+1)\Gamma(n_{i}+1)}
{\Gamma(n_{f}+2s_{f}+1)\Gamma(n_{i}+2s_{i}+1)}\right]^
{\frac{1}{2}} .
\end{eqnarray}
To derive an explicit expression for the integral (\ref{Sleta}) one can apply
the substitution $c\eta^{2}=x$ with $dx=2c\eta d\eta$, such that
\begin{eqnarray}
\eta^{l+1}d\eta =\frac{1}{2c^{1+l/2}}x^{l/2}dx.
\end{eqnarray}
Then Eq.~(\ref{Sleta}) reads as
\begin{eqnarray}
S_{l}(n_i,I_i;n_f,I_f)
=\frac{N_{n_i,n_f}(c,s_i,s_f)}{2c^{1+l/2}}\int_{0}^{\infty}
e^{-x}x^{s_{i}+s_{f}+\frac{l}{2}}L_{n_{f}}^{2s_{f}}(x)
L_{n_{i}}^{2s_{i}}(x)dx.
\label{Slx}
\end{eqnarray}
By using known formulas for integration of two generalized Laguerre polynomials
with different real ranks \cite{Prudnikov}, \cite{Wlagint} one obtains
(\ref{Slx}) in the following explicit form
\begin{eqnarray}
& &S_{l}(n_i,I_i;n_f,I_f)
\nonumber \\
&=&\frac{N_{n_i,n_f}(c,s_i,s_f)}{2c^{1+l/2}}
\frac{\Gamma (n_{f}+2s_{f}+1)}{\Gamma(1+2s_{f})}
\frac{\Gamma(n_{i}+s_{i}-s_{f}-\frac{l}{2})}{\Gamma(s_{i}-s_{f}-1)}
\frac{\Gamma(s_{i}+s_{f}+\frac{l}{2}+1)} {n_{i}!n_{f}!} \label{sgen} \\
&\times&
_{3}F_{2}\left(-n_{f},s_{i}+s_{f}+\frac{l}{2}+1,s_{f}-s_{i}+\frac{l}{2}+1;
2s_{f}+1,s_{f}-s_{i}+\frac{l}{2}+1-n_{i};1\right),
\nonumber
\end{eqnarray}
where $_{3}F_{2}$ denotes a generalized hypergeometric function \cite{3F2}. The
generalized hypergeometric function $_{3}F_{2}$ is calculated numerically
through a summation of its series representation for which a Fortran code is
available \cite{genhyp93}. It can be easily checked that if the first argument
of $_{3}F_{2}$ in (\ref{sgen}) is zero, $n_f=0$, one has $_{3}F_{2}=1$. In this
case Eq.~(\ref{sgen}) reduces to the following simpler expression
\begin{eqnarray}
S_{l}(n_i,I_i;0,I_f)
&=&\frac{1}{c^{\,l/2}}\frac{\Gamma(s_{i}+s_{f}+\frac{l}{2}+1)
\Gamma(n_{i}+s_{i}-s_{f}-\frac{l}{2})}
{\sqrt{n_{i}!\Gamma(2s_{f}+1)\Gamma(n_{i}+2s_{i}+1)}
\Gamma(s_{i}-s_{f}-\frac{l}{2})}.
\label{snf0}
\end{eqnarray}
This corresponds to a transition from a non-yrast to an yrast state. The
integrals for the yrast intraband transitions, Eqs. (50) and (51) in
\cite{b2b3mod}, are directly obtained from Eq.~(\ref{snf0}) when $n_i=0$.
Simple explicit forms of the $S_l$ integrals for interband and intraband
transitions in the particular cases up to $n=2$, which are of practical
interest, are given below
\begin{eqnarray}
& &S_{l}(1,I_i;1,I_f)\nonumber \\
&=& \frac{1}{c^{\,l/2}}\Bigl[(2s_{i}+1)(2s_{f}+1)-
(s_{i}+s_{f}-\frac{l}{2})(s_{i}+s_{f}+\frac{l}{2}+1)\Bigr] \nonumber \\
&\times&\frac{\Gamma(s_{i}+s_{f}+\frac{l}{2}+1)}
{\sqrt{\Gamma(2s_{i}+2)\Gamma(2s_{f}+2)}},
\end{eqnarray}
\begin{eqnarray}
& &S_{l}(2,I_i;1,I_f)\nonumber \\
&=& \frac{\sqrt{2}}{2c^{\,l/2}}\biggl\{2(s_{i}+1)(2s_{i}+1)(2s_{f}+1)-
(s_{i}+s_{f}+\frac{l}{2}+1)\biggr. \nonumber \\
&\times&\biggl. \Bigl[2(s_{i}+1)(2s_{i}+4s_{f}+3) -(s_{i}+s_{f}+\frac{l}{2}+2)
(3s_{i}+s_{f}-\frac{l}{2}+2)  \Bigr]   \biggr\} \nonumber \\
&\times&\frac{\Gamma(s_{i}+s_{f}+\frac{l}{2}+1)}
{\sqrt{\Gamma(2s_{i}+3)\Gamma(2s_{f}+2)}}.
\end{eqnarray}

\begin{eqnarray}
& &S_{l}(2,I_i;2,I_f)\nonumber \\
&=&\frac{1}{2c^{\,l/2}} \Biggl\{ 4(s_{i}+1)(2s_{i}+1)(s_{f}+1)(2s_{f}+1)\Biggr.
\nonumber \\
&-&(s_{i}+s_{f}+\frac{l}{2}+1)\biggl[ 16(s_{i}+1)(s_{f}+1)(s_{i}+s_{f}+1)
\biggr.
\nonumber \\
&-&(s_{i}+s_{f}+\frac{l}{2}+2)\Bigl\{2(s_{i}+1)(2s_{i}+1)+2(s_{f}+1)(2s_{f}+1)
+16(s_{i}+1)(s_{f}+1) \Bigr.
\nonumber \\
&-&\Biggl.\biggl.\Bigl.(s_{i}+s_{f}+\frac{l}{2}+3)(3s_{i}+3s_{f}-\frac{l}{2}+4)
\Bigr\}\biggr] \Biggr\}
\frac{\Gamma(s_{i}+s_{f}+\frac{l}{2}+1)}{\sqrt{\Gamma(2s_{i}+3)\Gamma(2s_{f}+3)}}.
\end{eqnarray}

\section*{\bf Appendix C: Explicit form of the integrals over $\phi$}

The integrals over the angular variable $\phi$, (\ref{Ilam}), with the relevant
parities $\pi_{i}$ and $\pi_{f}$ can be obtained in the following explicit
forms. For $\lambda =2$ the integral $I_{2}^{\pm\pm}$ with
$k_1=k_2=k=\mbox{odd}$ (++) or even $(--)$ is
\begin{eqnarray}
I_{2}^{\pm\pm}(k)
=\frac{2}{\pi}\mbox{Cat}+\frac{(-1)^{k+1}}{4k}\left[1+\frac{4}{\pi}
\sum_{m=1}^{2k-1}\frac{\sin (m\pi/2)}{m}\right],
\end{eqnarray}
where $\mbox{Cat}=\sum_{n=0}^{\infty} \frac{(-1)^n}{(2n+1)^{2}}\approx 0.915
965 594 177...$ is the Catalan constant. In the case of $k_1\neq k_2$, both odd
or even, the integral is
\begin{eqnarray}
I_{2}^{\pm\pm}(k_1,k_2)&=&
\frac{1}{2|k_2-k_1|}\left[1+\frac{4}{\pi}\sum_{m=1}^{|k_2-k_1|-1}\frac{\sin
(m\pi/2)}{m} \right]\\
\nonumber
&+&\frac{(-1)^{k_{1}+1}}{2(k_2+k_1)}\left[1+\frac{4}{\pi}\sum_{m=1}^{k_2+k_1-1}
\frac{\sin (m\pi/2)}{m} \right].
\end{eqnarray}
For $\lambda =3$ one has
\begin{eqnarray}
I_{3}^{+-}(k_1,k_2)=\frac{2k_2}{k_2^{2}-k_1^{2}}-
\frac{1}{\pi}\left[\frac{(-1)^{(k_2-k_1-1)/2}}
{(k_2-k_1)^2}+\frac{(-1)^{(k_2+k_1-1)/2}}{(k_2+k_1)^2}
\right],
\end{eqnarray}
where $ k_1= 1, 3, 5, \dots \ ,\ \   k_2= 2, 4, 6,
\dots$
For $\lambda =1$ the integral is obtained in the form of an infinite,
but reasonably converging series
\begin{eqnarray}
I_{1}^{+-}&=&\frac{1}{2\pi}\sum_{m=\pm1}^{\pm\infty}
\sum_{n=\pm1}^{\pm\infty} \sum_{\nu =\pm 1}
\frac{\mbox{sign}(-n)}{|mn|}\nonumber \\
&\times&\left[(1-\delta_{k_2+\nu k_1,-m-n})
\frac{\sin[(k_2+\nu k_1+m+n)\frac{\pi}{2}]}{(k_2+\nu k_1+m+n)}+
\frac{\pi}{2}\delta_{k_2+\nu k_1,-m-n}\right],
\end{eqnarray}
where $ k_1= 1, 3, 5, \dots \ ,\ \   k_2= 2, 4, 6,
\dots$

\newpage

\baselineskip=14.5pt plus 1pt minus 1pt

%%%%%%%%%%%%%%%%%%%%%%%%%%%%%%% Tables %%%%%%%%%%%%%%%%%%%%%%%%%%%%%%%%%%%%%

\begin{table}[h]
\caption{Parameters of the model fits.}
\begin{center}
\begin{tabular}{ccccccc}
\hline\hline
Nucl & $\omega$ [MeV/$\hbar$] & $b$ [$\hbar^{-2}$] & $d_0$ [$\hbar^{2}$]&
$c$ & $p$ & $e^{1}_{eff}$ [e] \\
\hline \\

$^{152}$Sm & 0.295 & 2.450 &  78.8 & 113.2 & 0.854 & 1.01  \\
% k: (1,8)(1,8)

$^{154}$Sm & 0.205 & 4.625 & 108.5 & 132.6 & 0.808 & 1.017 \\
% k: (1,12)(11,14)(7,14)

$^{154}$Gd & 0.306 & 2.948 & 114.7 & 113.4 & 0.777 & 1.048 \\
% k: (1,10)(1,8)(3,6)

$^{156}$Gd & 0.439 & 1.642 & 197.6 & 141.5 & 0.849 & 0.723 \\
% k: (1,8)(3,6)

$^{158}$Gd & 0.168 & 3.626 &  42.6 &  39.7 & 0.864 & 0.435 \\
% k: (3,12)(11,12)(11,14)

$^{236}$U  & 0.402 & 1.404 & 539.3 & 343.4 & 0.949 & 0.134 \\
% k: (1,8)(3,4)

$^{100}$Mo & 0.318 & 2.674 & 1.366 &  54.6 & 0.715 & 0.282 \\
% k: (1,6)(1,6)

& & & & & & \\

\hline\hline
\end{tabular}
\end{center}
\end{table}

\begin{table}[h]
\caption{Resulting mass parameters $B_2$ and $B_3$ (in $\hbar^{2}/$MeV),
Eq.~(\ref{Hqo}), and parameters of the model potential $C_2$ and $C_3$ (in
MeV), $d_2$ and $d_3$ (in $\hbar^{2}/$MeV), Eq.~(\ref{Ub2b3I}). The semi-axes
(sa) of the ellipsoidal potential bottom $\beta_{2}^{\mbox{\scriptsize sa}}$
and $\beta_{3}^{\mbox{\scriptsize sa}}$, Eq.~(\ref{semiax}), at angular
momentum $I=0$ are given in columns 8 and 9.}
\begin{center}
\begin{tabular}{ccccccccc}
\hline\hline
Nucl & $B_2$ & $B_3$ & $C_2$ & $C_3$ & $d_2$ & $d_3$
& $\beta_{2}^{\mbox{\scriptsize sa}}$
& $\beta_{3}^{\mbox{\scriptsize sa}}$ \\
\hline \\

$^{152}$Sm & 525 & 241 & 45.8 & 21.0 & 429 & 197 & 0.252 & 0.371 \\
% k: (1,8)(1,8)

$^{154}$Sm & 987 & 303 & 41.7 & 12.8 & 427 & 131 & 0.279 & 0.504 \\
% k: (1,12)(11,14)(7,14)

$^{154}$Gd & 613 & 127 & 57.6 & 11.9 & 416 & 86  & 0.263 & 0.578 \\
% k: (1,10)(1,8)(3,6)

$^{156}$Gd & 447 & 197 & 86.2 & 38.0 & 545 & 240 &  0.255 & 0.384 \\
% k: (1,8)(3,6)

$^{158}$Gd & 317 & 156 &  8.9 & 4.4  & 175 & 86 & 0.407 & 0.579 \\
% k: (3,12)(11,12)(11,14)

$^{236}$U  & 948 & 760 & 153 & 123 & 1351 & 1083 & 0.226 & 0.252 \\
% k: (1,8)(3,4)

$^{100}$Mo & 337 & 7 & 34.0 & 0.7  & 252 & 6 & 0.112 & 0.759 \\
% k: (1,6)(1,6)

& & & & & & & & \\

\hline\hline
\end{tabular}
\end{center}
\end{table}

\newpage

\tablecaption{Theoretical and experimental values of B(E1), B(E2) and B(E3)
transition probabilities in Weisskopf units (W.u.) for alternating-parity
spectra of several even-even nuclei. Notations: $g$ (ground-state band), $b1$
(first $\beta$-band), $b2$ (second $\beta$-band), $n1$ (first negative-parity
band), $n2$ (second negative-parity band), $n3$ (third negative-parity band).
The data are taken from \cite{nudat2} except for those for
B$(E3;3_{n1}^{-}\rightarrow  0_{g}^{+})$ transitions, which are taken from
\cite{K02}. The parity signs $(+)$ for the even and $(-)$ for the odd angular
momenta, respectively are omitted in the labels of the states to avoid
overloading of notations. The uncertainties (in parentheses) refer to the last
significant digits in the experimental data.}
\begin{center}

\tablefirsthead{\hline\hline
Mult & Transition & Th [W.u.] & Exp [W.u.]& Mult & Transition & Th [W.u.]
& Exp [W.u.] \\
\hline \\ }
\tablehead{\multicolumn{8}{l}{Table 3, {\em continued}} \\ \hline\hline
Mult & Transition & Th [W.u.] & Exp [W.u.]& Mult & Transition & Th [W.u.]
& Exp [W.u.] \\
\hline \\ }
\tabletail{\multicolumn{8}{r}{\em continues on next page} \\ \hline\hline}
\tablelasttail{\hline \hline}

\begin{supertabular}{cccccccccc}

\multicolumn{8}{c}{$^{152}$Sm} \\

& & & & & & & \\

$E2$ &$2_{g}$  $\rightarrow$   $0_{g}$ & 141  &144 (3)
&  $E2$  &$3_{n2}$ $\rightarrow$  $1_{n2}$&  52  &    \\

$E2$ &$4_{g}$  $\rightarrow$   $2_{g}$ & 210  &209 (3)
& $E2$  &$5_{n2}$ $\rightarrow$  $3_{n2}$&   63  &    \\

$E2$ &$6_{g}$  $\rightarrow$   $4_{g}$ & 248  &245 (5)
& $E3$  &$3_{n1}$ $\rightarrow$  $0_{g}$&   14  & 14 (2) \\

$E2$ &$8_{g}$  $\rightarrow$   $6_{g}$ & 284  &285 (14)
&$E3$  &$3_{n2}$  $\rightarrow$ $0_{b1}$&   10   &   \\

$E2$ &$10_{g}$ $\rightarrow$   $8_{g}$ & 322  &320 (3)
&$E3$  &$1_{n1}$ $\rightarrow$  $4_{g}$ &   69   &   \\

$E2$ &$12_{g}$ $\rightarrow$   $10_{g}$ & 363  &
&$E3$  &$1_{n2}$ $\rightarrow$  $4_{b1}$&   70   &   \\

$E1$ &$1_{n1}$ $\rightarrow$  $0_{g}$  &0.0041 &    0.0042 (4)
&$E2$  &$2_{b1}$ $\rightarrow$  $0_{g}$ &  1.26  &   0.92 (8) \\

$E1$ &$1_{n1}$ $\rightarrow$  $2_{g}$  &0.0088 &    0.0077 (7)
&$E2$  &$4_{b1}$ $\rightarrow$  $2_{g}$ &  0.2  &   0.7 (2) \\

$E1$ &$3_{n1}$ $\rightarrow$ $2_{g}$  &0.0056 &    0.0081 (16)
&$E2$  &$2_{b1}$ $\rightarrow$  $2_{g}$ &  4.6  &   5.5 (5) \\

$E1$ &$3_{n1}$ $\rightarrow$ $4_{g}$  &0.0087 &    0.0082 (16)
&$E2$  &$4_{b1}$ $\rightarrow$  $4_{g}$ &  4.2  &   5.4 (13)\\

$E1$ &$1_{n2}$ $\rightarrow$ $0_{b1}$ &0.0041 &
&$E2$  &$2_{b1}$ $\rightarrow$  $4_{g}$ &  27.4  &   19.2 (18) \\

$E1$ &$1_{n2}$ $\rightarrow$ $2_{b1}$ &0.0095 &
&$E2$  &$4_{b1}$ $\rightarrow$  $6_{g}$ &  35  &    4 (2) \\

$E2$  &$2_{b1}$ $\rightarrow$  $0_{b1}$ & 160 &    107 (27)
&$E2$  &$0_{b1}$ $\rightarrow$  $2_{g}$&   30    &   \\

$E2$  &$4_{b1}$ $\rightarrow$  $2_{b1}$ & 232 &    204  (38)
&$E1$  &$1_{n1}$ $\rightarrow$  $2_{b1}$&  0.00402&  0.00013 (4) \\

$E2$  &$3_{n1}$ $\rightarrow$  $1_{n1}$&  47  &
&$E1$ &$1_{n1}$  $\rightarrow$  $0_{b1}$&   0.0023 &  \\

$E2$  &$5_{n1}$ $\rightarrow$  $3_{n1}$&  58  &
&$E1$ &$1_{n2}$  $\rightarrow$    $0_{g}$&   0.00006&  \\
& & & & & & & \\

\multicolumn{8}{c}{$^{154}$Sm} \\

& & & & & & & \\

$E2$ &$2_{g}$  $\rightarrow$   $0_{g}$ & 168  &176 (1)
&$E2$  &$5_{n2}$ $\rightarrow$  $3_{n2}$&  82  &    \\

$E2$ &$4_{g}$  $\rightarrow$   $2_{g}$ & 247  &245 (6)
&$E2$  &$3_{n3}$ $\rightarrow$  $1_{n3}$&  72  &    \\

$E2$ &$6_{g}$  $\rightarrow$   $4_{g}$ & 287  &289 (8)
&$E3$  &$3_{n1}$ $\rightarrow$  $0_{g} $&   10  & 10 (2) \\

$E2$ &$8_{g}$  $\rightarrow$   $6_{g}$ & 322  &319 (17)
&$E3$  &$1_{n1}$ $\rightarrow$  $4_{g}$ &   50   &  \\

$E2$ &$10_{g}$ $\rightarrow$   $8_{g}$ & 358  &314 (16)
&$E3$  &$3_{n2}$  $\rightarrow$ $0_{b1}$&   77   &   \\

$E2$ &$12_{g}$ $\rightarrow$   $10_{g}$ & 398 &282 (19)
&$E3$  &$1_{n2}$ $\rightarrow$  $4_{b1}$&   381  &   \\

$E1$ &$1_{n1}$ $\rightarrow$  $0_{g}$  &0.0051 &  0.0058 (4)
&$E3$  &$3_{n3}$  $\rightarrow$ $0_{b2}$&   6    &   \\

$E1$ &$1_{n1}$ $\rightarrow$  $2_{g}$  &0.0110 &  0.0113 (7)
&$E3$  &$1_{n3}$ $\rightarrow$  $4_{b2}$&   62   &   \\

$E1$ &$3_{n1}$ $\rightarrow$  $2_{g}$  &0.0069 &  0.0080 (11)
&$E2$  &$0_{b1}$ $\rightarrow$  $2_{g}$ &   1  & 12 (3) \\

$E1$ &$3_{n1}$ $\rightarrow$  $4_{g}$  &0.0106 &  0.0092 (13)
&$E2$  &$2_{b1}$ $\rightarrow$  $0_{g}$ & 0.36 &   $<$0.58 \\

$E1$ &$1_{n2}$ $\rightarrow$ $0_{b1}$ &0.0109 &
&$E2$  &$2_{b1}$ $\rightarrow$  $2_{g}$ & 0.39 &   $<$1.3 \\

$E1$ &$1_{n2}$ $\rightarrow$ $2_{b1}$ &0.0231 &
&$E2$  &$2_{b1}$ $\rightarrow$  $4_{g}$ & 0.27&   $<$2.4 \\

$E1$ &$1_{n3}$ $\rightarrow$ $0_{b2}$ &0.0044 &
&$E2$  &$0_{b2}$ $\rightarrow$  $2_{g}$ & $5\times 10^{-6}$ & \\

$E1$ &$1_{n3}$ $\rightarrow$ $2_{b2}$ &0.0109 &
&$E2$  &$0_{b2}$ $\rightarrow$  $2_{b1}$ & 16 &  \\

$E2$  &$2_{b1}$ $\rightarrow$  $0_{b1}$ & 65 &
&$E1$  &$0_{b1}$ $\rightarrow$  $1_{n1}$ &  0.0005  &  \\

$E2$  &$4_{b1}$ $\rightarrow$  $2_{b1}$ & 93 &
&$E1$  &$1_{n2}$ $\rightarrow$  $0_{g}$  &  0.0005  &  \\

$E2$  &$2_{b2}$ $\rightarrow$  $0_{b2}$ & 68 &
&$E1$  &$1_{n2}$ $\rightarrow$  $0_{b2}$&  0.0058   &  \\

$E2$  &$4_{b2}$ $\rightarrow$  $2_{b2}$ & 97 &
&$E1$  &$1_{n3}$  $\rightarrow$  $0_{b1}$& $3\times 10^{-7}$ & \\

$E2$  &$3_{n1}$ $\rightarrow$  $1_{n1}$ & 60&
&$E1$  &$1_{n3}$  $\rightarrow$   $0_{g}$& $8\times 10^{-5}$& \\

$E2$  &$5_{n1}$ $\rightarrow$  $3_{n1}$ & 72&
&$E3$  &$3_{n2}$  $\rightarrow$   $0_{g}$ &   1.7&  \\

$E2$  &$3_{n2}$ $\rightarrow$  $1_{n2}$&  69  &
&$E3$  &$3_{n3}$ $\rightarrow$   $0_{g}$  &  0.4 &  \\

& & & & & & & \\

\multicolumn{8}{c}{$^{154}$Gd} \\

& & & & & & & \\

$E2$ &$2_{g}$  $\rightarrow$   $0_{g}$ & 160  &157 (1)
&$E2$   &$5_{n2}$ $\rightarrow$  $3_{n2}$&  64   &    \\

$E2$ &$4_{g}$  $\rightarrow$   $2_{g}$ & 235  &245 (9)
&$E2$   &$3_{n3}$ $\rightarrow$  $1_{n3}$&  51   &     \\

$E2$ &$6_{g}$  $\rightarrow$   $4_{g}$ & 273  &285 (15)
&$E3$  &$3_{n1}$ $\rightarrow$  $0_{g} $&   21   & 21 (5) \\

$E2$ &$8_{g}$  $\rightarrow$   $6_{g}$ & 306  &312 (17)
&$E3$  &$3_{n2}$  $\rightarrow$ $0_{b1}$&   32   &   \\

$E2$ &$10_{g}$ $\rightarrow$   $8_{g}$ & 340  &360 (4)
&$E3$  &$3_{n3}$  $\rightarrow$ $0_{b2}$&  144   &  \\

$E2$ &$12_{g}$ $\rightarrow$   $10_{g}$& 377  &
&$E3$  &$1_{n1}$ $\rightarrow$  $4_{g}$ &  102   &   \\

$E1$ &$1_{n1}$ $\rightarrow$  $0_{g}$  &0.0102 &    0.0436
&$E3$  &$1_{n2}$ $\rightarrow$  $4_{b1}$&  179   &   \\

$E1$ &$1_{n1}$ $\rightarrow$  $2_{g}$  &0.0216 &    0.0485
&$E3$  &$1_{n3}$ $\rightarrow$  $4_{b2}$&  708   &   \\

$E1$ &$3_{n1}$ $\rightarrow$  $2_{g}$  &0.0137 &
&$E2$  &$0_{b1}$ $\rightarrow$  $2_{g}$ &  25    & 52 (8) \\

$E1$ &$3_{n1}$ $\rightarrow$  $4_{g}$  &0.0207 &
&$E2$  &$2_{b1}$ $\rightarrow$  $0_{g}$ &  1.23  &  0.86 (7) \\

$E1$ &$1_{n2}$ $\rightarrow$ $0_{b1}$  &0.0152 &
&$E2$  &$2_{b1}$ $\rightarrow$  $4_{g}$ & 22.6  &  19.6 (16) \\

$E1$ &$1_{n2}$ $\rightarrow$ $2_{b1}$  &0.0333 &
&$E2$  &$0_{b2}$ $\rightarrow$  $2_{g}$ &  0.0553 &   \\

$E1$ &$1_{n3}$ $\rightarrow$ $0_{b2}$  &0.0333 &
&$E2$  &$0_{b2}$ $\rightarrow$  $2_{b1}$ & 14     &  \\

$E1$ &$1_{n3}$ $\rightarrow$ $2_{b2}$  &0.0706 &
&$E1$  &$1_{n1}$ $\rightarrow$  $0_{b1}$ &  0.0054 & 0.0057  \\

$E2$  &$2_{b1}$ $\rightarrow$  $0_{b1}$ & 177  &   97 (10)
&$E1$  &$1_{n1}$ $\rightarrow$  $2_{b1}$ &  0.0099  & 0.0064 \\

$E2$  &$4_{b1}$ $\rightarrow$  $2_{b1}$ & 256  &
&$E1$  &$1_{n2}$ $\rightarrow$  $0_{g}$  & 2$\times 10^{-5}$ & \\

$E2$  &$2_{b2}$ $\rightarrow$  $0_{b2}$ & 85   &
&$E1$  &$1_{n2}$ $\rightarrow$  $0_{b2}$ &  0.0094   &  \\

$E2$  &$4_{b2}$ $\rightarrow$  $2_{b2}$ & 122  &
&$E1$  &$1_{n3}$  $\rightarrow$ $0_{b1}$ & 0.00023   &  \\

$E2$  &$3_{n1}$ $\rightarrow$  $1_{n1}$ & 55   &
&$E1$  &$1_{n3}$  $\rightarrow$ $0_{g}$  & 2$\times 10^{-6}$& \\

$E2$  &$5_{n1}$ $\rightarrow$  $3_{n1}$ & 67   &
&$E3$  &$3_{n2}$  $\rightarrow$ $0_{g}$  &   1.8     &    \\

$E2$  &$3_{n2}$ $\rightarrow$  $1_{n2}$ & 54   &
&$E3$  &$3_{n3}$  $\rightarrow$ $0_{g}$  &   0.05&   \\

& & & & & & & \\

\multicolumn{8}{c}{$^{156}$Gd} \\

& & & & & & & \\

$E2$ &$2_{g}$  $\rightarrow$   $0_{g}$ & 150  &187 (5)
&  $E2$  &$3_{n2}$ $\rightarrow$  $1_{n2}$&  44  &  \\

$E2$ &$4_{g}$  $\rightarrow$   $2_{g}$ & 219  &263 (5)
& $E2$   &$5_{n2}$ $\rightarrow$  $3_{n2}$&  53  &   \\

$E2$ &$6_{g}$  $\rightarrow$   $4_{g}$ & 249  &295 (8)
& $E3$  &$3_{n1}$ $\rightarrow$  $0_{g}$&  16.9  &  16.9 (7) \\

$E2$ &$8_{g}$  $\rightarrow$   $6_{g}$ & 273  &320 (14)
& $E3$  &$3_{n2}$  $\rightarrow$ $0_{b1}$&  64   & \\

$E2$ &$10_{g}$ $\rightarrow$   $8_{g}$ & 296  &314 (14)
&$E3$  &$1_{n1}$ $\rightarrow$  $4_{g}$ &   73   & \\

$E2$ &$12_{g}$ $\rightarrow$   $10_{g}$ & 321 &300 (3)
&$E3$  &$1_{n2}$ $\rightarrow$  $4_{b1}$&   282  & \\

$E1$ &$1_{n1}$ $\rightarrow$  $0_{g}$  &0.0006  &0.0019 (14)
&$E2$  &$0_{b1}$ $\rightarrow$  $2_{g}$ &   5    &  8 (4) \\

$E1$ &$1_{n1}$ $\rightarrow$  $2_{g}$  &0.0013  &0.0025 (18)
&$E2$  &$2_{b1}$ $\rightarrow$  $0_{g}$ &  0.32  &  0.63 (6) \\

$E1$ &$3_{n1}$ $\rightarrow$  $2_{g}$  &0.00083 &0.00098 (21)
&$E2$  &$4_{b1}$ $\rightarrow$  $2_{g}$ &  0.1   &   1.3  (7) \\

$E1$ &$3_{n1}$ $\rightarrow$  $4_{g}$  &0.0012  &0.00077 (16)
&$E2$  &$4_{b1}$ $\rightarrow$  $6_{g}$ &  5.6   &   2.1 (11) \\

$E1$ &$1_{n2}$ $\rightarrow$  $0_{b1}$ &0.0013  &
&$E2$  &$2_{b1}$ $\rightarrow$  $4_{g}$ &  4.3 &   4.1  (4) \\

$E1$ &$1_{n2}$ $\rightarrow$  $2_{b1}$ &0.0026  &0.0005 (3)
&$E1$ &$1_{n1}$ $\rightarrow$  $0_{b1}$ &0.0002  &0.0004 (3) \\

$E1$ &$3_{n2}$ $\rightarrow$  $2_{b1}$ &0.0016  &
&$E1$  &$1_{n2}$ $\rightarrow$  $0_{g}$& $6\times 10^{-6}$& 0.0019 (7)\\

$E2$  &$2_{b1}$ $\rightarrow$  $0_{b1}$ & 74    &  52 (23)
&$E1$  &$1_{n2}$ $\rightarrow$  $2_{g}$& $2\times 10^{-5}$& 0.0043 (15)\\

$E2$  &$4_{b1}$ $\rightarrow$  $2_{b1}$ & 107   & 280 (15)
&$E1$ &$3_{n2}$  $\rightarrow$  $2_{g}$& $5\times 10^{-6}$& 0.0019 (14)\\

$E2$  &$6_{b1}$ $\rightarrow$  $4_{b1}$ & 120   &
&$E1$ &$3_{n2}$  $\rightarrow$ $4_{g}$&  $2\times 10^{-5}$&  0.0031 (4)\\

$E2$  &$3_{n1}$ $\rightarrow$  $1_{n1}$&  46    &
&$E3$ &$3_{n2}$  $\rightarrow$    $0_{g}$&   0.21&  \\

$E2$  &$5_{n1}$ $\rightarrow$  $3_{n1}$&  56  &    \\

& & & & & & & \\
\multicolumn{8}{c}{$^{158}$Gd} \\

& & & & & & & \\

$E2$ &$2_{g}$  $\rightarrow$   $0_{g}$ & 181  &198 (6)
&$E2$  &$0_{b1}$ $\rightarrow$  $2_{g}$ & 8.7619  & 1.1652  \\

$E2$ &$4_{g}$  $\rightarrow$   $2_{g}$ & 274  &289 (5)
&$E2$  &$2_{b1}$ $\rightarrow$  $0_{g}$ & 2.36 &   0.31 (4) \\

$E2$ &$6_{g}$  $\rightarrow$   $4_{g}$ & 332  &
&$E2$  &$2_{b1}$ $\rightarrow$  $2_{g}$ & 2.913&   0.079 (14)\\

$E2$ &$8_{g}$  $\rightarrow$   $6_{g}$ & 393  &330 (3)
&$E2$  &$4_{b1}$ $\rightarrow$  $4_{g}$ & 2.40 &   0.37  \\

$E2$ &$10_{g}$ $\rightarrow$   $8_{g}$ & 460  &340 (3)
&$E2$  &$2_{b1}$ $\rightarrow$  $4_{g}$ & 2.96 &   1.39 (15) \\

$E2$ &$12_{g}$ $\rightarrow$   $10_{g}$ & 532 &310 (3)
&$E2$  &$0_{b2}$ $\rightarrow$  $2_{g}$ & 1.86 & 2.09   \\

$E1$ &$1_{n1}$ $\rightarrow$  $0_{g}$ &0.0001 & 9.8443$\times
10^{-5}$(4)
&$E2$  &$2_{b2}$ $\rightarrow$  $0_{g}$ & 0.68 &  0.37 (4)  \\

$E1$ &$1_{n1}$ $\rightarrow$  $2_{g}$  &2.5$\times 10^{-4}$ &
9.6515$\times 10^{-5}(6)$
&$E2$  &$2_{b2}$ $\rightarrow$  $4_{g}$ & 0.43 &  0.38 (6) \\

$E1$ &$3_{n1}$ $\rightarrow$  $2_{g}$  &0.00015&  0.00033 (10)
&$E2$  &$4_{b1}$ $\rightarrow$  $2_{g}$ & 3.75 &  1.32 \\

$E1$ &$3_{n1}$ $\rightarrow$  $4_{g}$  &0.00028&  0.00029 (8)
&$E2$  &$4_{b1}$ $\rightarrow$  $6_{g}$ & 1.30 &  3.16  \\

$E1$ &$5_{n1}$ $\rightarrow$  $4_{g}$ &2.02$\times 10^{-4}$
&7.4324$\times 10^{-4}$(13)
&$E2$  &$0_{b2}$ $\rightarrow$  $2_{b1}$ & 57 &   \\

$E1$ &$5_{n1}$ $\rightarrow$  $6_{g}$ &3.62$\times 10^{-4}$
&5.8691$\times 10^{-4}$(8)
&$E1$  &$0_{b1}$ $\rightarrow$  $1_{n1}$&2.7$\times 10^{-5}$
&3.314$\times 10^{-6}$ \\

$E1$ &$3_{n2}$ $\rightarrow$  $2_{b1}$ &0.00011 &$>$ 0.00035
&$E1$  &$2_{b1}$ $\rightarrow$  $1_{n1}$&8.3$\times 10^{-6}$
&6.4$\times 10^{-5}$(8)\\

$E1$ &$1_{n2}$ $\rightarrow$ $0_{b1}$  &8.02$\times 10^{-5}$ &
&$E1$  &$2_{b1}$ $\rightarrow$  $3_{n1}$&2$\times 10^{-5}$
&1.89$\times 10^{-4}$(24) \\

$E1$ &$1_{n2}$ $\rightarrow$ $2_{b1}$  &0.0002 &
&$E1$  &$1_{n2}$ $\rightarrow$  $2_{g}$ & 4$\times 10^{-5}$
&  0.0064  \\

$E1$ &$1_{n3}$ $\rightarrow$ $0_{b2}$  &0.0004 &
&$E1$  &$1_{n2}$ $\rightarrow$  $0_{g}$ & 2$\times 10^{-5}$
& 0.0035 (12) \\

$E1$ &$1_{n3}$ $\rightarrow$ $2_{b2}$  &0.0009 &
&$E1$  &$3_{n2}$ $\rightarrow$  $2_{g}$ & 3$\times 10^{-5}$
& $>$0.0011 \\

$E1$ &$3_{n3}$ $\rightarrow$ $2_{b2}$  &0.0005 &
&$E1$  &$3_{n2}$ $\rightarrow$  $4_{g}$ & 3$\times 10^{-5}$
& $>$0.0015 \\

$E2$  &$2_{b1}$ $\rightarrow$  $0_{b1}$ & 200 &
&$E1$  &$0_{b2}$ $\rightarrow$  $1_{n1}$ & 2$\times 10^{-7}$
& 5.7831$\times 10^{-5}$  \\

$E2$  &$4_{b1}$ $\rightarrow$  $2_{b1}$ & 288 & 455
&$E1$  &$2_{b2}$ $\rightarrow$  $1_{n1}$ & 2$\times 10^{-8}$
& 2.7$\times 10^{-6}$(19) \\

$E2$  &$2_{b2}$ $\rightarrow$  $0_{b2}$ & 217 &
&$E1$  &$2_{b2}$ $\rightarrow$  $3_{n1}$& 2$\times 10^{-7}$
& 3.7$\times 10^{-5}$(5)  \\

$E2$  &$4_{b2}$ $\rightarrow$  $2_{b2}$ & 308 &
&$E1$  &$0_{b2}$  $\rightarrow$ $1_{n2}$& 6$\times 10^{-5}$
&6.02$\times 10^{-4}$   \\

$E2$  &$3_{n1}$ $\rightarrow$  $1_{n1}$ & 185 &
&$E1$  &$2_{b2}$  $\rightarrow$ $1_{n2}$& 1.8$\times 10^{-5}$
&1.50$\times 10^{-4}$(21)  \\

$E2$  &$5_{n1}$ $\rightarrow$  $3_{n1}$ & 227& 369 (6)
&$E1$  &$2_{b2}$  $\rightarrow$ $3_{n2}$& 4.2$\times 10^{-5}$
& 2.40$\times 10^{-4}$(5) \\

$E2$  &$3_{n2}$ $\rightarrow$  $1_{n2}$ & 200& $>$ 1600
&$E1$  &$4_{b1}$  $\rightarrow$ $3_{n1}$ &7.7$\times 10^{-6}$
& 4.63$\times 10^{-4}$   \\

$E2$  &$5_{n2}$ $\rightarrow$  $3_{n2}$ & 240&
&$E1$  &$4_{b1}$  $\rightarrow$ $5_{n1}$& 2.1$\times 10^{-5}$
& 6.12$\times 10^{-4}$   \\

$E2$  &$3_{n3}$ $\rightarrow$  $1_{n3}$& 241 &
&$E1$  &$1_{n2}$ $\rightarrow$  $0_{b2}$ & 2$\times 10^{-5}$ & \\

$E3$  &$3_{n1}$ $\rightarrow$  $0_{g} $&   11.9  & 11.9 (7)
&$E1$  &$1_{n3}$ $\rightarrow$  $0_{b1}$ & 3$\times 10^{-6}$ & \\

$E3$  &$1_{n1}$ $\rightarrow$  $4_{g}$ &   81   &
&$E1$  &$1_{n3}$ $\rightarrow$  $0_{g}$ & 0.00001 &  \\

$E3$  &$3_{n2}$ $\rightarrow$  $0_{b1}$ &   519   &
&$E3$  &$3_{n2}$ $\rightarrow$  $0_{g}$ &   5   &  \\

$E3$  &$3_{n3}$  $\rightarrow$ $0_{b2}$&   102  &
&$E3$  &$3_{n3}$ $\rightarrow$  $0_{g}$ &   2  &  \\

& & & & & & & \\

\multicolumn{8}{c}{$^{236}$U} \\

& & & & & & & \\

$E2$ &$2_{g}$  $\rightarrow$   $0_{g}$ &  237  &250 (10)
&$E2$  &$2_{b1}$ $\rightarrow$  $0_{b1}$ & 112 &  \\

$E2$ &$4_{g}$  $\rightarrow$   $2_{g}$ &  342  &357 (23)
&$E2$  &$4_{b1}$ $\rightarrow$  $2_{b1}$ & 160 &  \\

$E2$ &$6_{g}$  $\rightarrow$   $4_{g}$ &  382  &385 (22)
&$E2$  &$3_{n1}$ $\rightarrow$  $1_{n1}$ &  68 &  \\

$E2$ &$8_{g}$  $\rightarrow$   $6_{g}$ &  408  &390 (4)
&$E2$  &$5_{n1}$ $\rightarrow$  $3_{n1}$&  80  &  \\

$E2$ &$10_{g}$ $\rightarrow$   $8_{g}$ &  429  &360 (4)
&$E2$  &$7_{n1}$ $\rightarrow$  $5_{n1}$&  87  &  \\

$E2$ &$12_{g}$ $\rightarrow$   $10_{g}$ & 450  &410 (7)
&$E2$  &$3_{n2}$ $\rightarrow$  $1_{n2}$&  54  &  \\

$E2$ &$14_{g}$ $\rightarrow$   $12_{g}$ & 471  &450 (5)
&$E2$  &$5_{n2}$ $\rightarrow$  $3_{n2}$&  64  &  \\

$E2$ &$16_{g}$ $\rightarrow$   $14_{g}$ & 493  &380 (4)
&$E3$  &$1_{n1}$ $\rightarrow$  $4_{g}$ &  62  & 62 (9) \\

$E2$ &$18_{g}$ $\rightarrow$   $16_{g}$ & 516  &490 (5)
&$E3$  &$3_{n1}$ $\rightarrow$  $0_{g}$ &  15  & 23 (3) \\

$E2$ &$20_{g}$ $\rightarrow$   $18_{g}$ & 539  &510 (8)
&$E3$  &$1_{n2}$ $\rightarrow$  $4_{b1}$&  695   &  \\

$E2$ &$22_{g}$ $\rightarrow$   $20_{g}$ & 564  &520 (12)
&$E3$  &$3_{n2}$ $\rightarrow$  $0_{b1}$&  172   &   \\

$E2$ &$24_{g}$ $\rightarrow$   $22_{g}$ & 590  &670 (13)
&$E2$  &$0_{b1}$ $\rightarrow$  $2_{g}$ &    6   &  \\

$E2$ &$26_{g}$ $\rightarrow$   $24_{g}$ & 617  &670 (19)
&$E2$  &$2_{b1}$ $\rightarrow$  $0_{g}$ &    0.66 &  \\

$E2$ &$28_{g}$ $\rightarrow$   $26_{g}$ & 645  &1100 (5)
&$E2$  &$4_{b1}$ $\rightarrow$  $2_{g}$ &    0.59 &  \\

$E1$ &$1_{n1}$ $\rightarrow$  $0_{g}$  &2.7$ \times 10^{-8}$
& 2.7$ \times 10^{-8}$(4)
&$E2$  &$2_{b1}$ $\rightarrow$  $4_{g}$ &    4    &  \\

$E1$ &$1_{n1}$ $\rightarrow$  $2_{g}$  &5.5$ \times 10^{-8}$&
&$E1$  &$0_{b1}$ $\rightarrow$  $1_{n1}$ &  1.2$ \times 10^{-8}$ \\

$E1$ &$3_{n1}$ $\rightarrow$ $2_{g}$   &3.5$ \times 10^{-8}$&
&$E1$  &$2_{b1}$ $\rightarrow$  $1_{n1}$ &  4.6$ \times 10^{-9}$  & \\

$E1$ &$3_{n1}$ $\rightarrow$ $4_{g}$   &4.8$ \times 10^{-8}$&
&$E1$  &$1_{n2}$ $\rightarrow$  $0_{g}$  &  1.6$ \times 10^{-9}$  & \\

$E1$ &$1_{n2}$ $\rightarrow$ $0_{b1}$  &2.0$ \times 10^{-8}$   &
&$E3$  &$3_{b2}$ $\rightarrow$    $0_{g}$ &  0.14  &    \\

$E1$ &$1_{n2}$ $\rightarrow$ $2_{b1}$  &4.0$ \times 10^{-8}$  &  \\

& & & & & & & \\

\multicolumn{8}{c}{$^{100}$Mo} \\

& & & & & & & \\

$E2$ &$2_{g}$  $\rightarrow$   $0_{g}$ & 22.7 &37.0 (7)
&$E2$  &$3_{n1}$ $\rightarrow$  $1_{n1}$ & 16 & \\

$E2$ &$4_{g}$  $\rightarrow$   $2_{g}$ & 50    &69  (4)
&$E2$  &$5_{n1}$ $\rightarrow$  $3_{n1}$ & 21 & \\

$E2$ &$6_{g}$  $\rightarrow$   $4_{g}$ & 84    &94  (14)
&$E2$  &$7_{n1}$ $\rightarrow$  $5_{n1}$ & 26 & \\

$E2$ &$8_{g}$  $\rightarrow$   $6_{g}$ & 120   &123 (18)
&$E2$  &$3_{n2}$ $\rightarrow$  $1_{n2} $& 18 & \\

$E2$ &$10_{g}$ $\rightarrow$   $8_{g}$ & 156  &
&$E2$  &$5_{n2}$ $\rightarrow$  $3_{n2} $& 22 & \\

$E1$ &$1_{n1}$ $\rightarrow$  $0_{g}$  &2$\times 10^{-6}$  &
&$E3$  &$3_{n1}$ $\rightarrow$  $0_{g}  $& 34 & 34 (3)\\

$E1$ &$1_{n1}$ $\rightarrow$  $2_{g}$  &1$\times 10^{-5}$&
&$E3$  &$3_{n2}$  $\rightarrow$ $0_{b1}$&   5  & \\

$E1$ &$3_{n1}$ $\rightarrow$  $2_{g}$  &7$\times 10^{-6}$&2.7$
\times 10^{-6}$(9)
&$E3$  &$1_{n1}$ $\rightarrow$  $4_{g}$&  899  & \\

$E1$ &$3_{n1}$ $\rightarrow$  $4_{g}$  &2$\times 10^{-5}$&
&$E2$  &$0_{b1}$ $\rightarrow$  $2_{g}$ &  72  & 92 (4) \\

$E1$ &$1_{n2}$ $\rightarrow$ $0_{b1}$  &2$\times 10^{-7}$&
&$E2$  &$2_{b1}$ $\rightarrow$  $0_{g}$ &  0.5 & 0.62 (5) \\

$E1$ &$1_{n2}$ $\rightarrow$ $2_{b1}$  & 1$\times 10^{-5}$ &
&$E2$  &$4_{b1}$ $\rightarrow$  $2_{g}$ & 3  &   \\

$E1$ &$3_{n2}$ $\rightarrow$ $2_{b1}$  &3$\times 10^{-6}$&
&$E1$  &$1_{n1}$ $\rightarrow$  $0_{b1}$ & $8\times 10^{-6}$ & \\

$E1$ &$3_{n2}$ $\rightarrow$ $4_{b1}$  &3$\times 10^{-5}$&
&$E1$  &$1_{n1}$ $\rightarrow$  $2_{b1}$ & $2\times10^{-5}$  & \\

$E2$ &$2_{b1}$ $\rightarrow$  $0_{b1}$ & 25.4&   5.5  (8)
&$E1$  &$3_{n1}$ $\rightarrow$  $2_{b1}$ &  1.4$\times 10^{-5}$  &
2.5$\times 10^{-5}$(8)  \\

$E2$  &$4_{b1}$ $\rightarrow$  $2_{b1}$ & 45 &
&$E1$  &$1_{n2}$ $\rightarrow$  $0_{g}$  & 5$\times 10^{-7}$&  \\

$E2$  &$6_{b1}$ $\rightarrow$  $4_{b1}$ & 75 &
&$E3$  &$3_{n2}$ $\rightarrow$  $0_{g}$&  22   &  \\

\end{supertabular}
\end{center}

\newpage
%%%%%%%%%%%%%%%%%%%%%%%%%%%%%%% Figures %%%%%%%%%%%%%%%%%%%%%%%%%%%%%%%%%%%%%

\begin{figure}%[t]
\centerline{ {\epsfxsize=18.cm\epsfbox{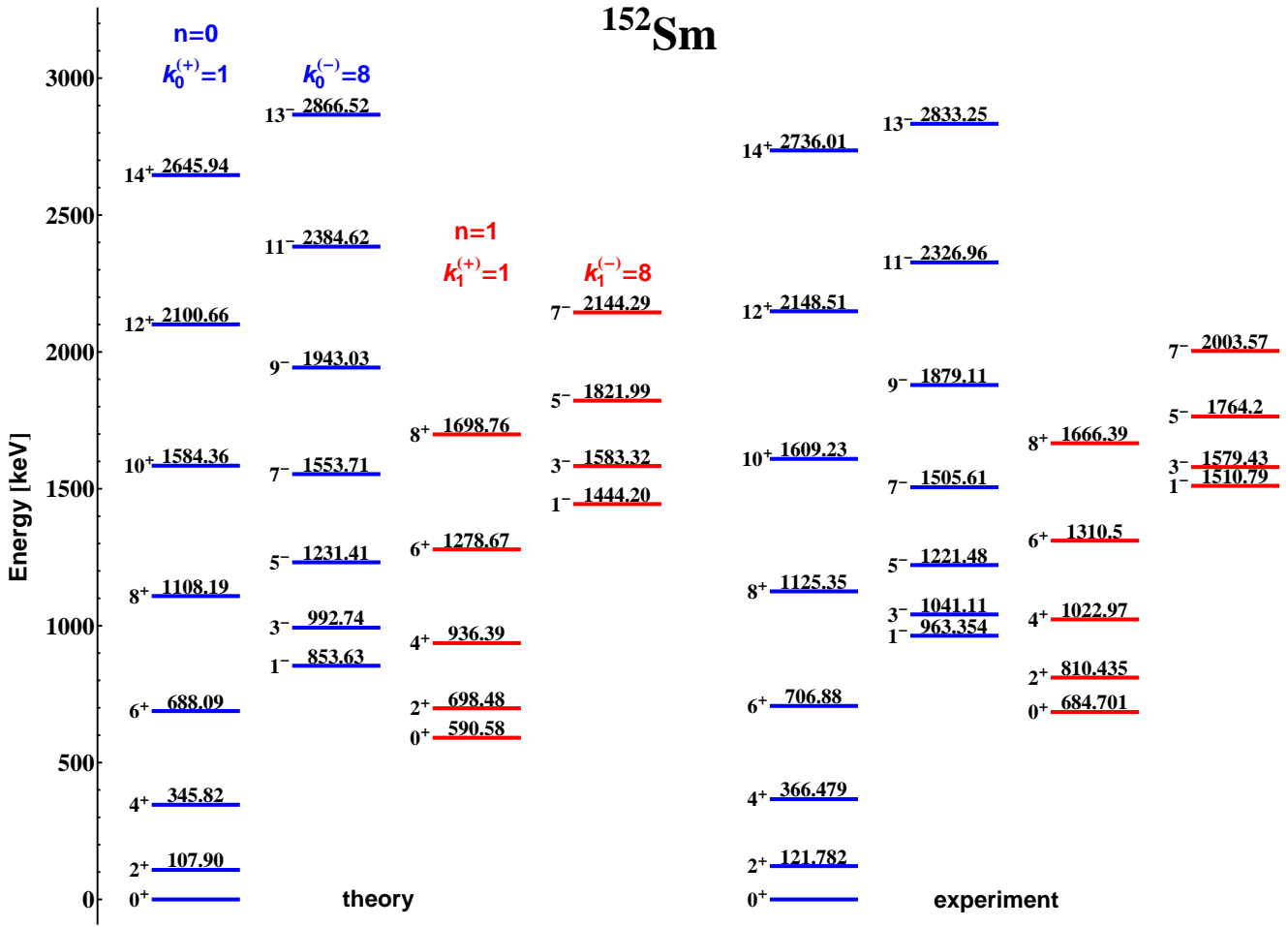}}}
\caption{(Color online) Theoretical and experimental alternating-parity bands
in $^{152}$Sm. Data from \cite{ensdf}. The oscillation quantum numbers $n$,
$k_n^{(+)}$ and $k_n^{(-)}$ are given above the theoretical bands.}
\end{figure}

\begin{figure}%[t]
\centerline{ {\epsfxsize=18.cm\epsfbox{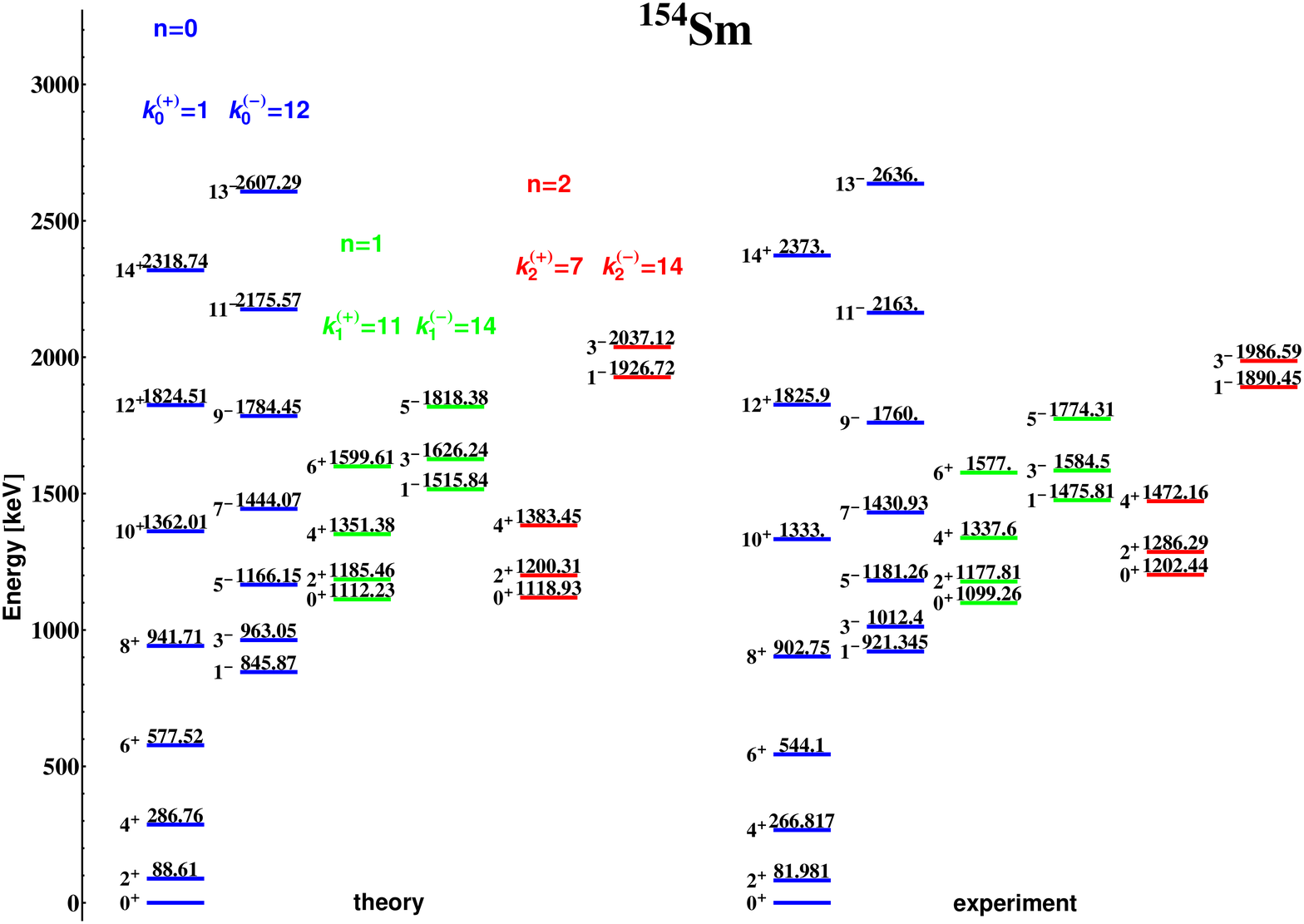}}}
\caption{(Color online) The same as in Fig.~1, but for $^{154}$Sm.}
\end{figure}

\begin{figure}%[t]
\centerline{ {\epsfxsize=18.cm\epsfbox{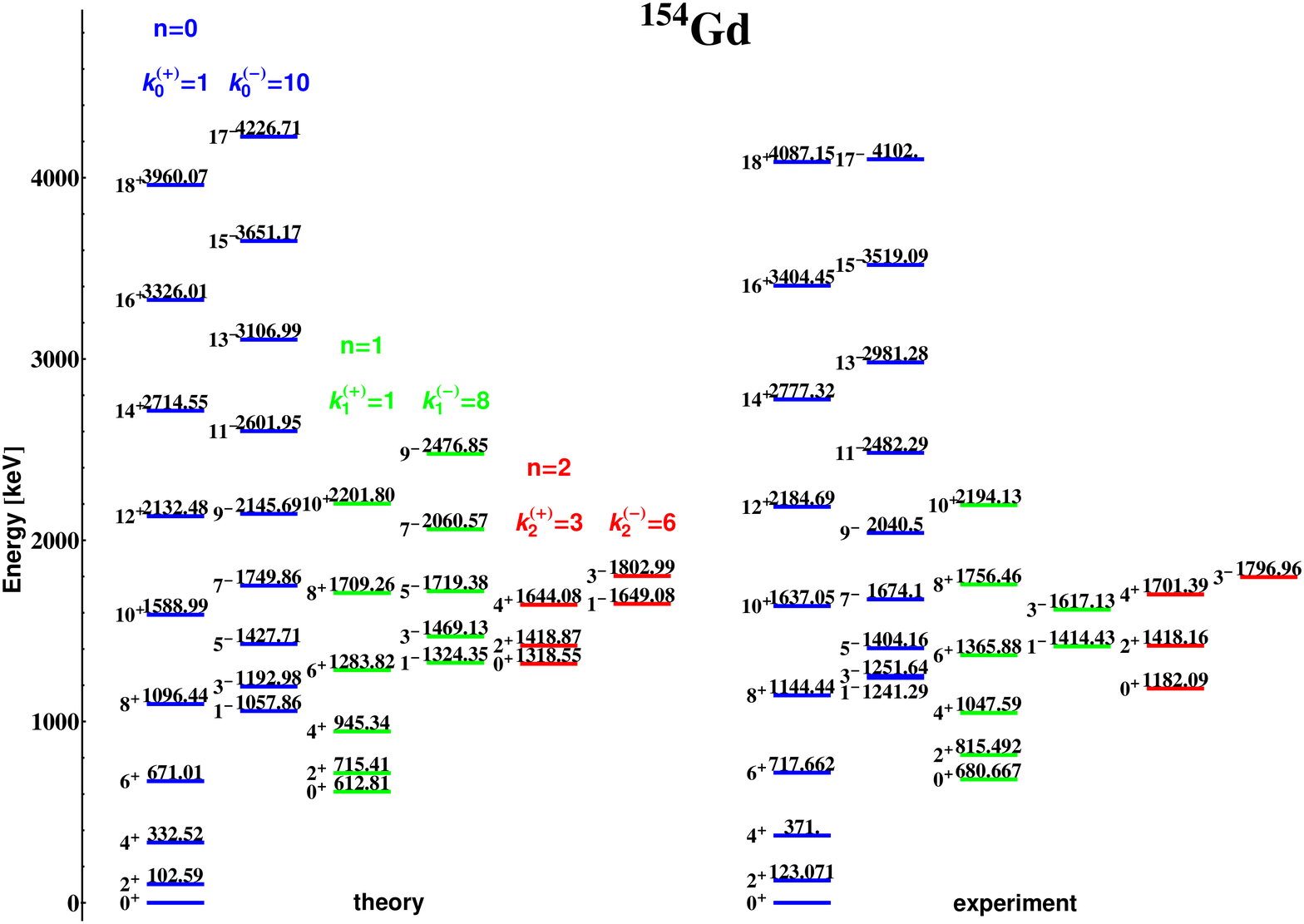}}}
\caption{(Color online) The same as in Fig.~1, but for $^{154}$Gd.}
\end{figure}

\begin{figure}%[t]
\centerline{ {\epsfxsize=18.cm\epsfbox{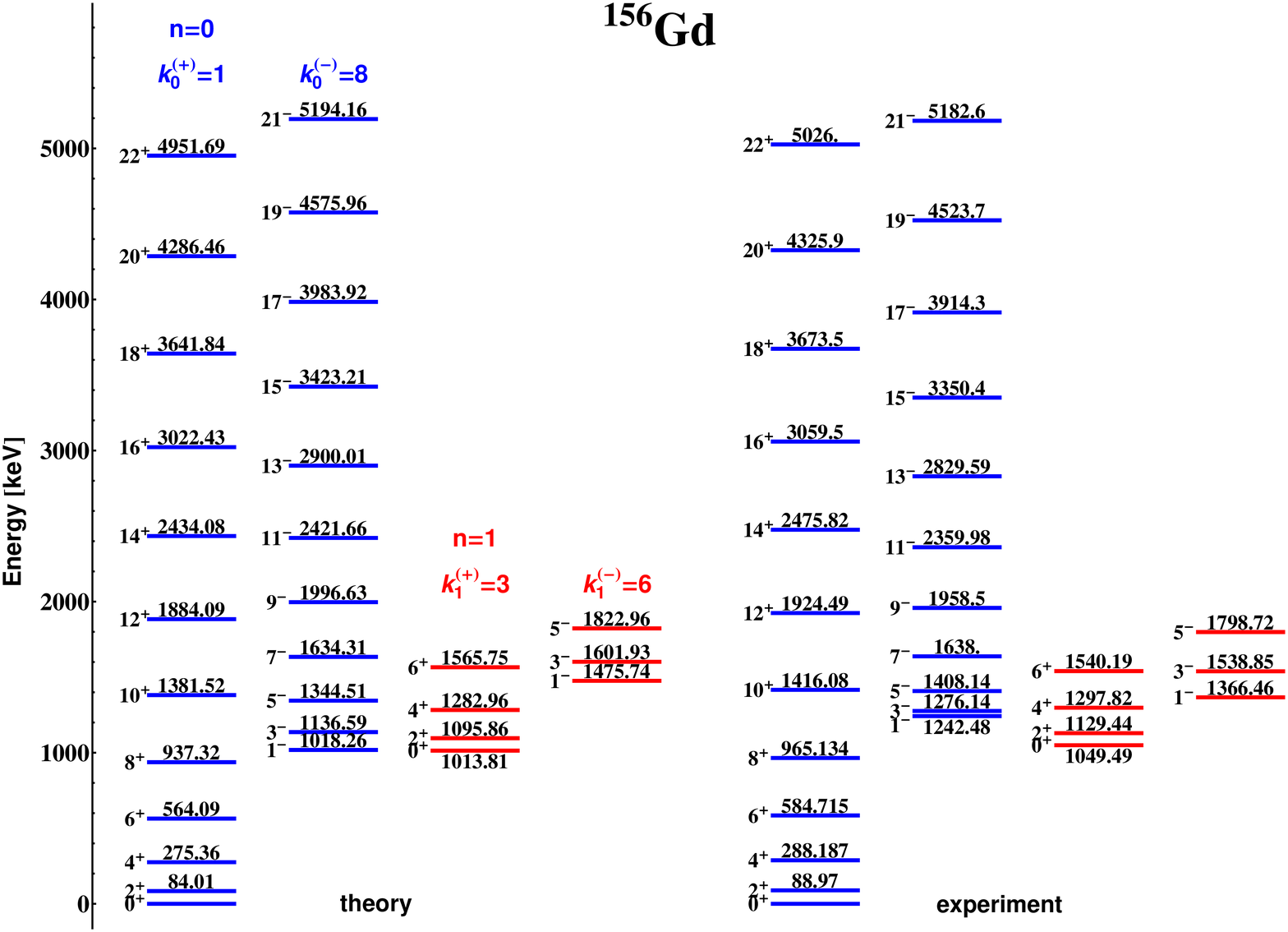}}}
\caption{(Color online) The same as in Fig.~1, but for $^{156}$Gd.}
\end{figure}

\begin{figure}%[t]
\centerline{ {\epsfxsize=18.cm\epsfbox{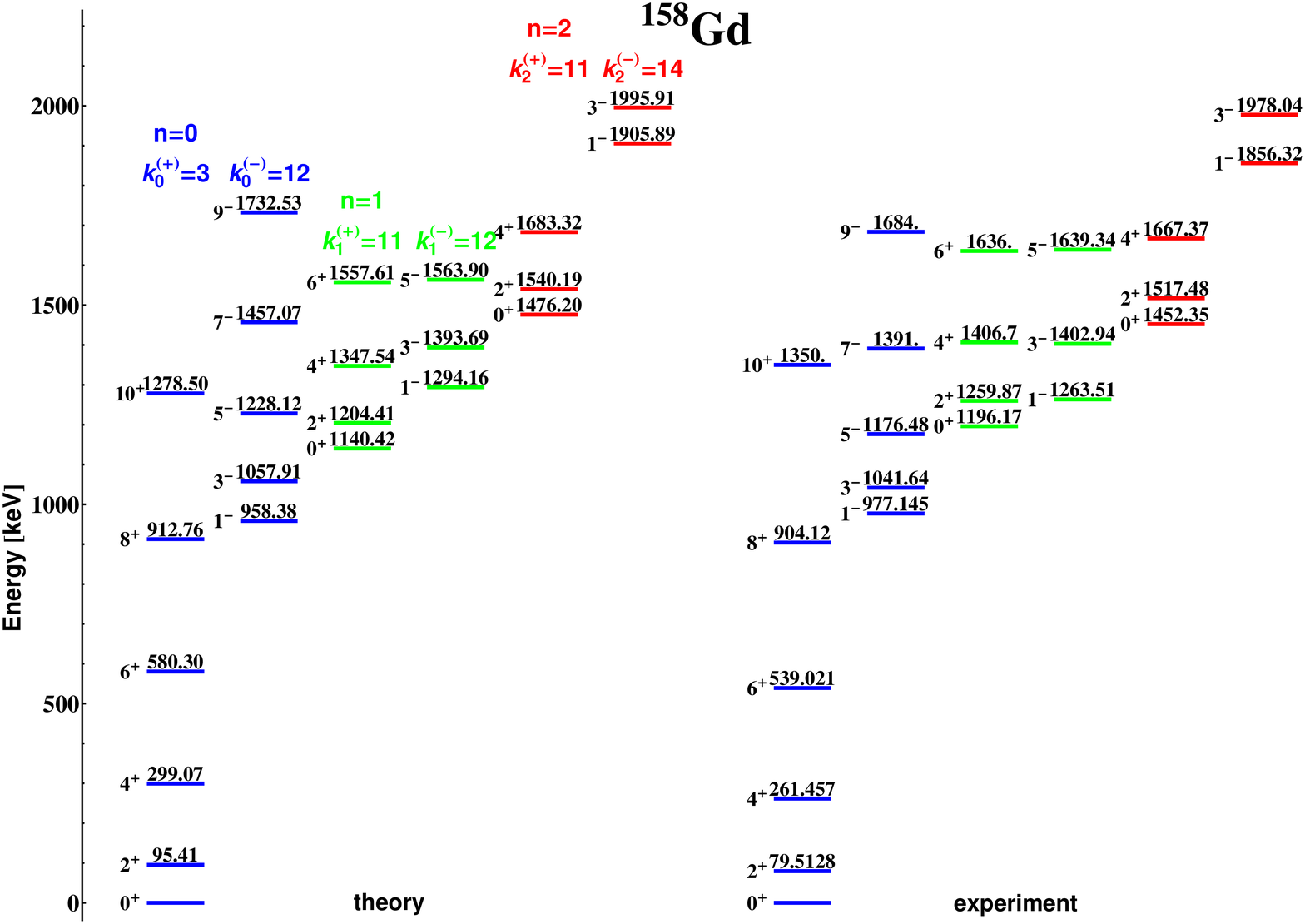}}}
\caption{(Color online) The same as in Fig.~1, but for $^{158}$Gd.}
\end{figure}

\begin{figure}%[t]
\centerline{ {\epsfxsize=18.cm\epsfbox{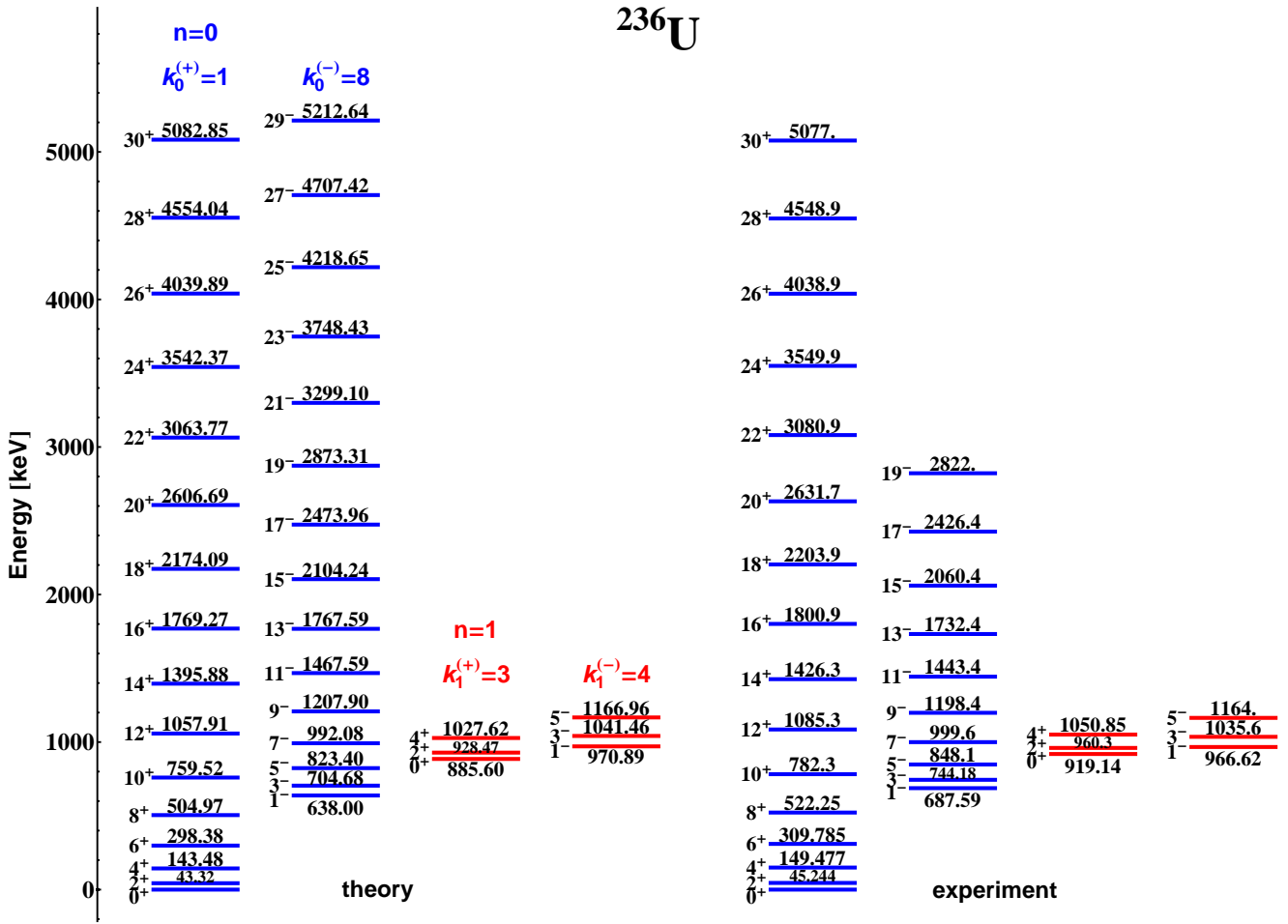}}}
\caption{(Color online) The same as in Fig.~1, but for $^{236}$U.}
\end{figure}

\begin{figure}%[t]
\centerline{ {\epsfxsize=18.cm\epsfbox{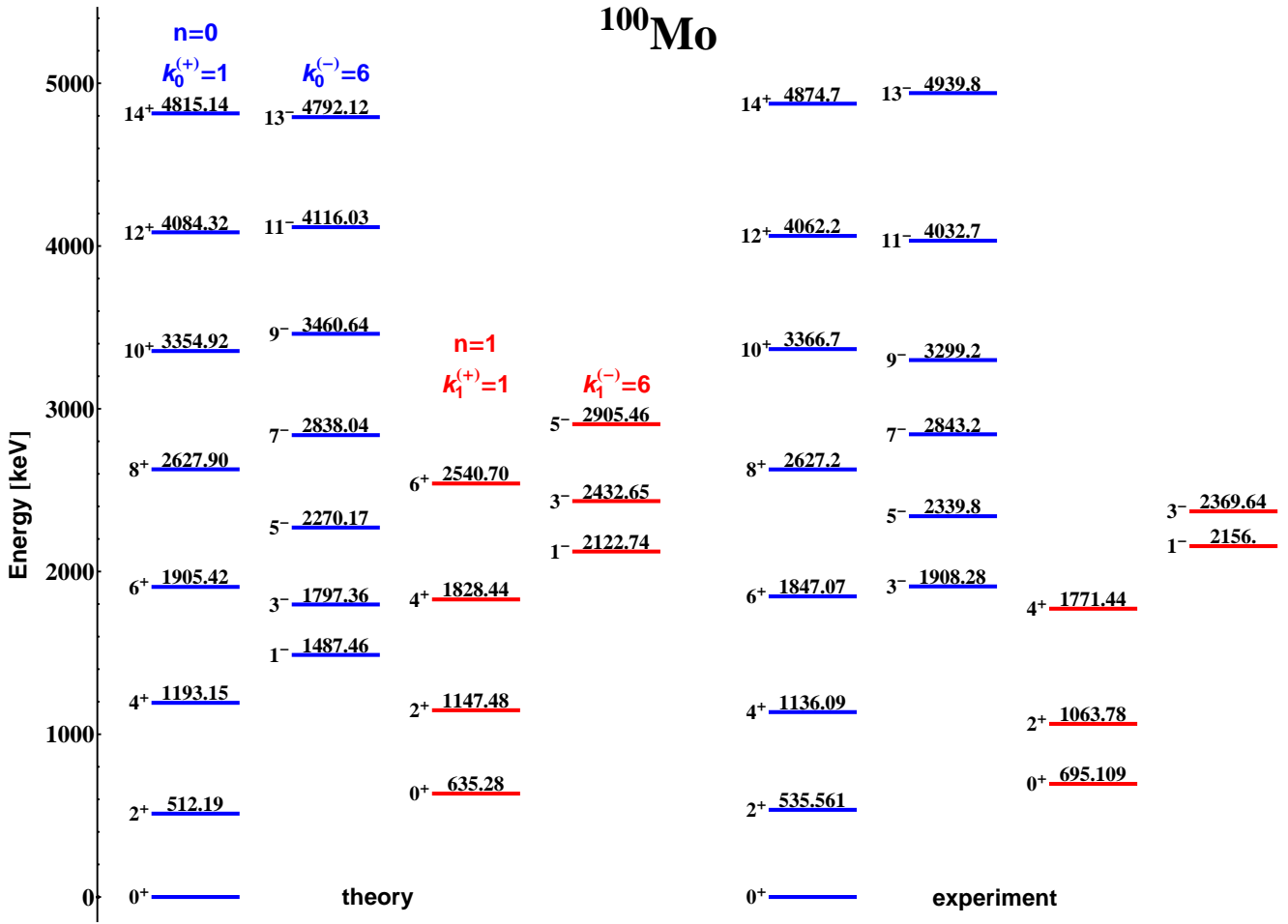}}}
\caption{(Color online) The same as in Fig.~1, but for $^{100}$Mo.}
\end{figure}

\ \

\newpage
\begin{figure}%[t]
\centerline{ {\epsfxsize=8.cm\epsfbox{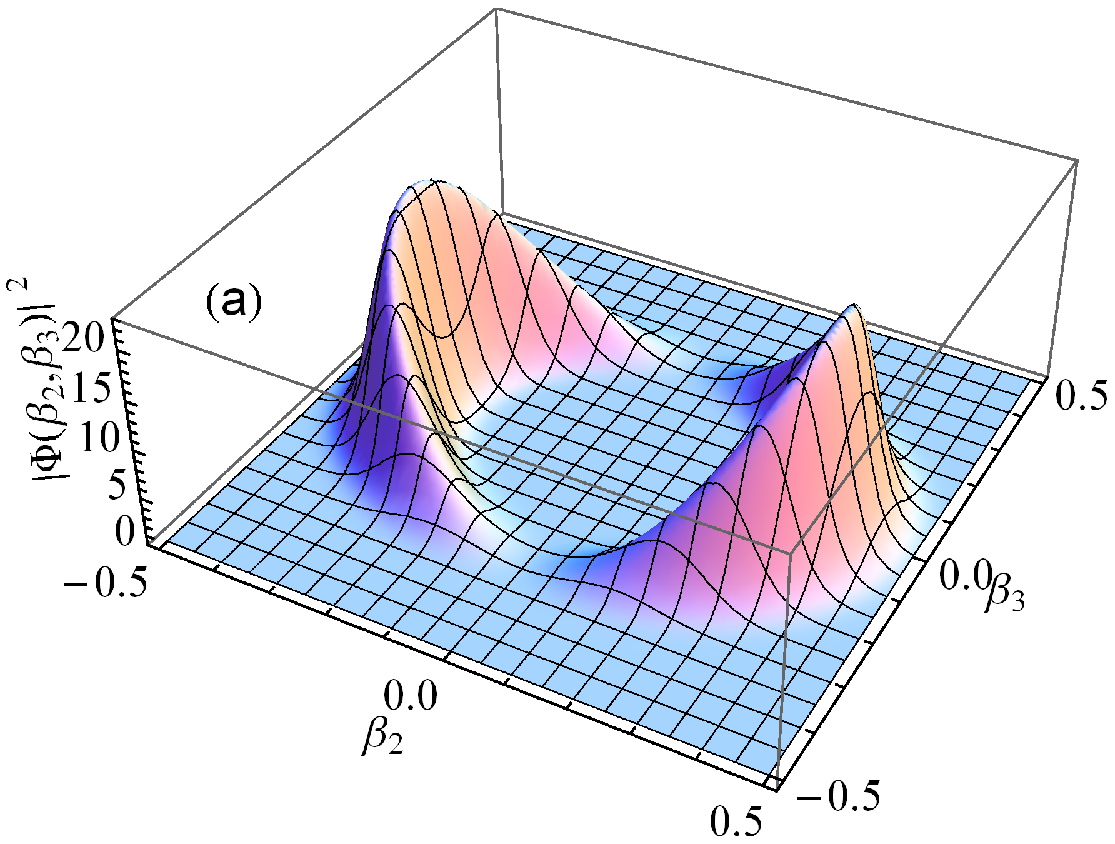}}\ \
{\epsfxsize=8.cm\epsfbox{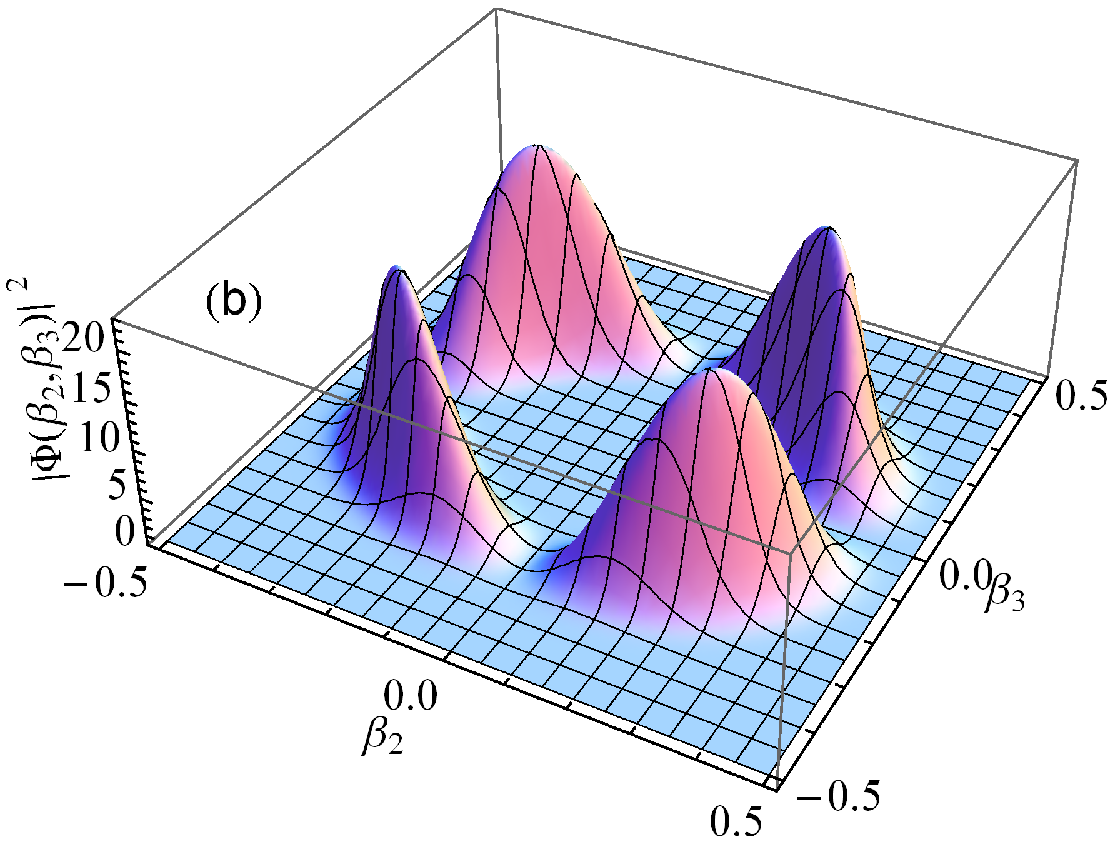}}  }
\caption{(Color online) Density distribution
$\rho_{nkI}(\beta_{2},\beta_{3})=|\Phi^{\pi}_{nkI} (\beta_{2},\beta_{3})|^{2}$
for: (a) $k=1$, $I=2$ and (b) $k=2$, $I=1$ at $n=0$ with schematic
parameters (see the text). The model space corresponds to the
$\beta_{2} >0$ half-plane.}
\end{figure}

\begin{figure}%[t]
\centerline{ {\epsfxsize=8.cm\epsfbox{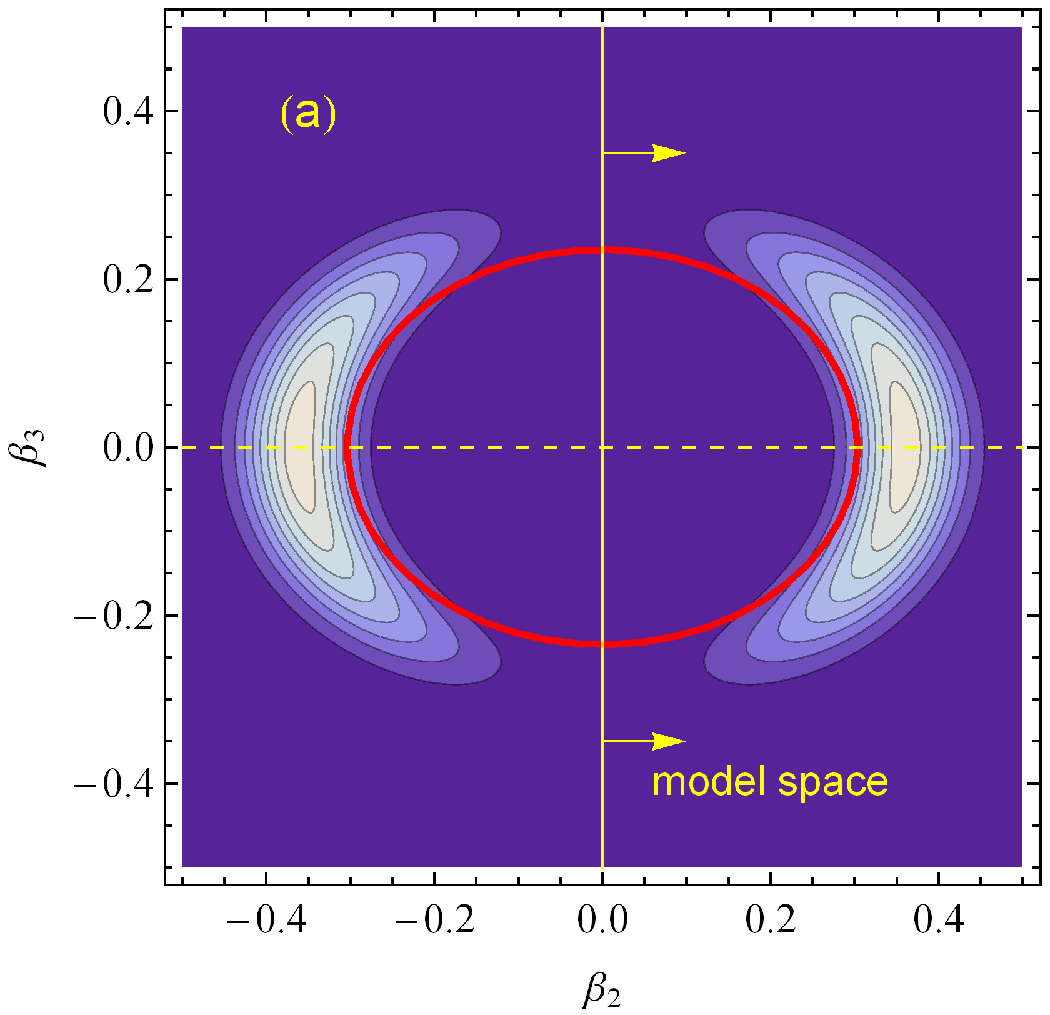}}\ \
{\epsfxsize=8.cm\epsfbox{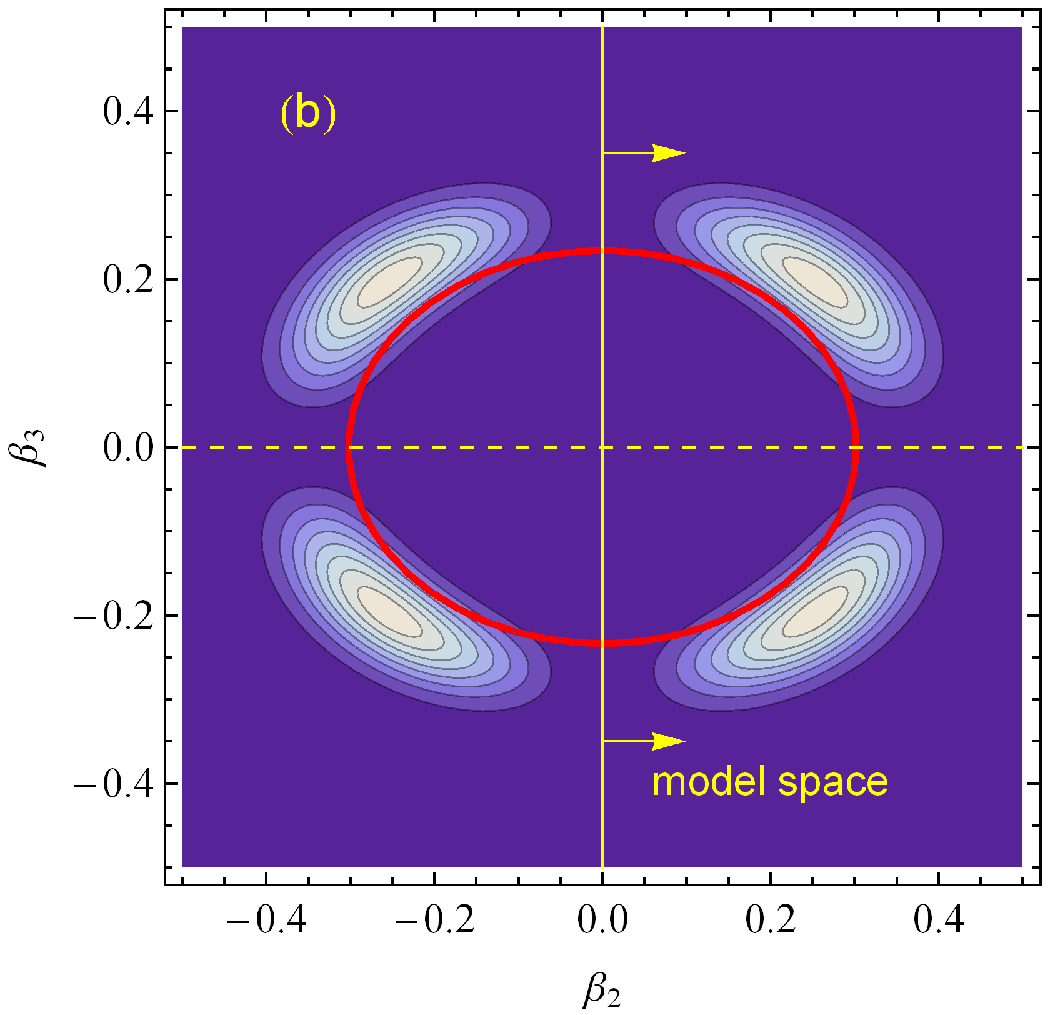}}}
\centerline{ {\epsfxsize=8.cm\epsfbox{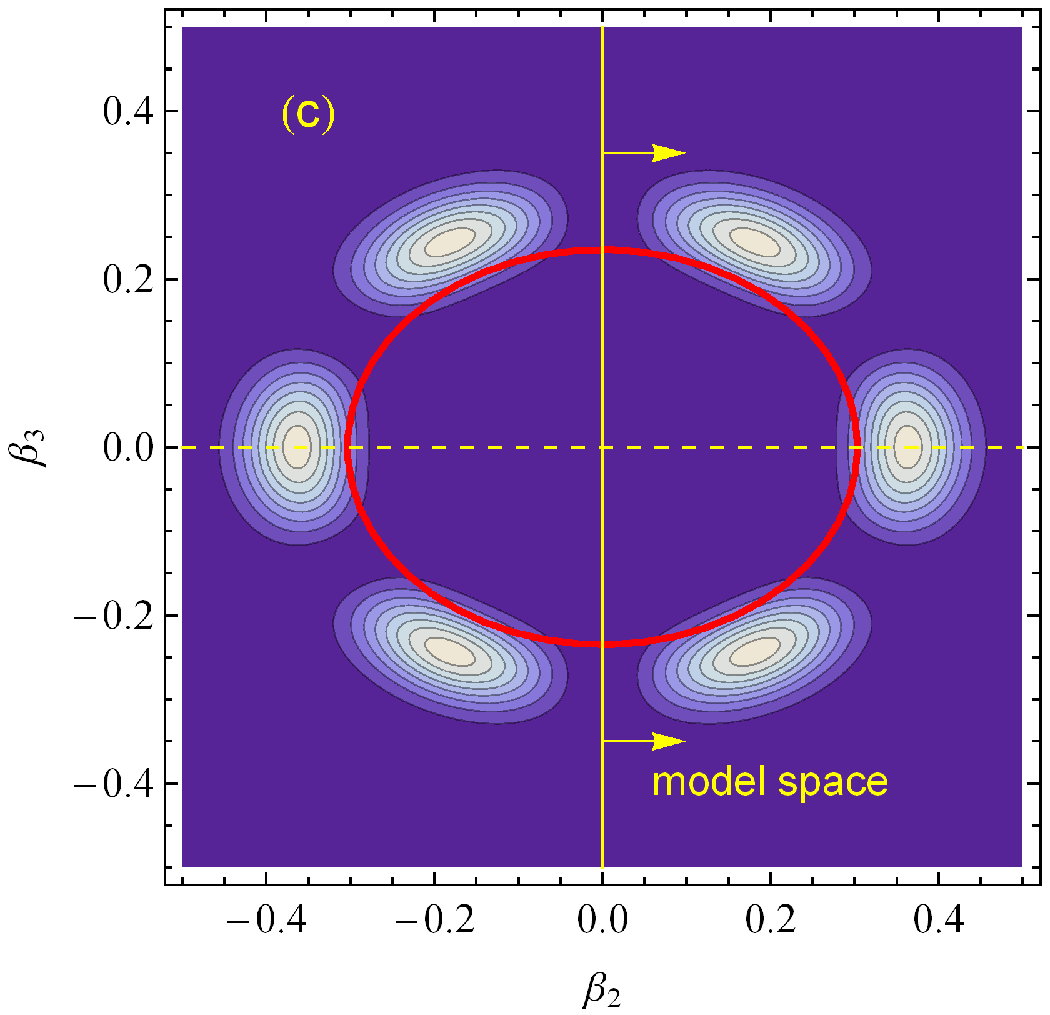}}\ \
{\epsfxsize=8.cm\epsfbox{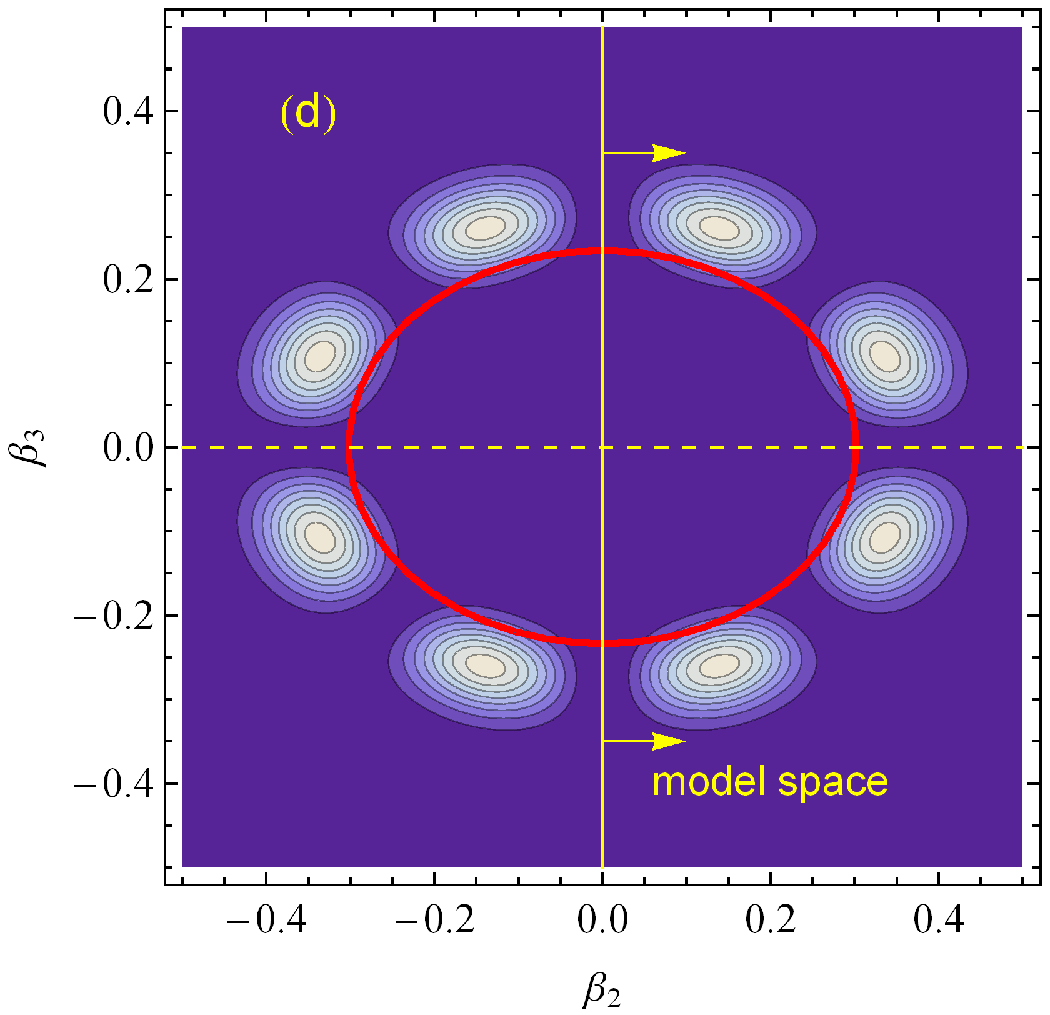}}}
\caption{(Color online) Contour plots of the density distribution
$\rho_{nkI}(\beta_{2},\beta_{3})$
for: (a) $k=1$, $I=2$, (b) $k=2$, $I=1$, (c) $k=3$, $I=2$
and (d) $k=4$, $I=1$ at $n=0$ with the schematic parameters
(see the text). The ellipsoidal curves outline the potential
bottom. The model space corresponds
to the $\beta_{2} >0$ half-plane.}
\end{figure}

\end{document}